\documentclass[twocolumn,prl,aps,10pt]{revtex4-2}

\usepackage{graphicx}
\usepackage{dcolumn}
\usepackage{bm}
\usepackage{multirow}
\usepackage{xcolor}
\usepackage[normalem]{ulem}
\usepackage{amsmath}
\usepackage{gensymb}
\usepackage[colorlinks=true]{hyperref}
\usepackage{booktabs}
\usepackage{threeparttable}

\begin{document}


\title{Twist-tuned exchange and hysteresis in a bilayer van der Waals magnet}

\author{Priyanka Mondal$^{1}$}
\author{Sonu Verma$^{2}$}
\author{Wenze Lan$^{1}$}
\author{Lukas Krelle$^{1}$}
\author{Ryan Tan$^{1}$}
\author{Regine von Klitzing$^{1}$}
\author{Kenji Watanabe$^{3}$}
\author{Takashi Taniguchi$^{4}$}
\author{Kseniia Mosina$^{5}$}
\author{Zdenek Sofer$^{5}$}
\author{Akashdeep Kamra$^{2}$}
\email{akashdeep.kamra@rptu.de}
\author{Bernhard Urbaszek$^1$}
\email{bernhard.urbaszek@pkm.tu-darmstadt.de}

\affiliation{\small$^1$Institute for Condensed Matter Physics, TU Darmstadt, Hochschulstraße 6-8, D-64289 Darmstadt, Germany
}
\affiliation{\small$^2$Department of Physics and Research Center OPTIMAS, Rheinland-Pf\"alzische Technische Universit\"at Kaiserslautern-Landau, 67663 Kaiserslautern, Germany}
\affiliation{\small$^3$Research Center for Electronic and Optical Materials,National Institute for Material Science,1-1 Namiki, Tsukuba 305-0044,Japan}
\affiliation{\small$^4$Research Center for Materials Nanoarchitectonics,National Institute for Material Science,1-1 Namiki, Tsukuba 305-0044,Japan}
\affiliation{\small$^5$Department of Inorganic Chemistry, University of Chemistry and Technology Prague, Technicka 5, 166 28 Prague 6, Czech Republic}


\maketitle

\textbf{Moir\'e superlattices in twisted bilayers enable profound reconstructions of the electronic bandstructure, giving rise to correlated states with remarkable tunability~\cite{cao2018unconventional,bistritzer2011moire,xia2025superconductivity}. Extending this paradigm to van der Waals magnets, twisting creates spatially varying interlayer exchange interactions that stabilize emergent spin textures~\cite{doi:10.1073/pnas.2000347117} and the coexistence of ferromagnetic and antiferromagnetic domains~\cite{song2021direct,xu2022coexisting,cheng2023electrically}. 
Here, we demonstrate the emergence of robust magnetic hysteresis in bilayer CrSBr upon twisting by an angle of $\sim$3°. This is observed as the corresponding hysteretic evolution of the exciton energy, that directly correlates with the bilayer magnetic state~\cite{wilson2021interlayer}, in magnetic field dependent photoluminescence measurements. A two-sublattice model captures this behavior, attributing it to the twist-induced reduction of interlayer exchange that stabilizes both parallel and antiparallel spin configurations across a broad field range. Comparison with experiment enables quantitative extraction of the effective exchange strength. Remarkably, the system exhibits coherent averaging across the moiré supercell, yielding an effective monodomain response characterized by switching into the antiferromagnetic state, rather than forming spin textures or fragmented domains. Spatially resolved measurements further uncover local variations in hysteresis loops, consistent with position-dependent modulation of the average exchange interaction. Our results establish twist engineering as a powerful route to programmable magnetic memories in two-dimensional magnets, harnessing the robustness of antiferromagnetic order.}

For two-dimensional (2D) quantum materials, twist engineering has emerged as a powerful technique to control correlated phenomena by deliberately misaligning atomic layers\cite{cao2018unconventional,carr2017twistronics,liao2020precise,choi2021twist,chen2024twist,inbar2023quantum,seiler2022quantum}. In bilayer graphene and, more recently, bilayer WSe$_2$, twisting modifies the electronic band structure, giving rise to correlated insulator phases and superconductivity\cite{bistritzer2011moire,xia2025superconductivity}. Here, the spatial coherence of electrons over the larger moir\'e supercell is essential for the emergence of a new bandstructure. Employing this approach in van der Waals (vdW) magnets\cite{huang2017layer,park20252d} opens new opportunities for spatially programmable magnetism, particularly in systems with layered spin order or antiferromagnetic coupling\cite{song2021direct,cheng2023electrically,xiao2021magnetization,xu2022coexisting,klein2022control,boix2025programmable,liu2024twisted}.

Chromium Sulfur Bromide (CrSBr) is a recently identified vdW magnet that combines strong excitonic features with layered antiferromagnetic (AFM) order\cite{telford2020layered,lee2021magnetic,bae2022exciton,Ziebel2024}. As an A-type antiferromagnet, CrSBr hosts ferromagnetically aligned spins within each layer, coupled antiferromagnetically between layers\cite{goser1990magnetic,lee2021magnetic,lopez2022dynamic}. A unique feature of CrSBr is its unusually strong intralayer ferromagnetic exchange that correlates with its high N\'eel temperature~\cite{Ziebel2024,Scheie2022}. Beyond magnetism, CrSBr exhibits strong coupling between lattice, electronic, and excitonic degrees of freedom, including exciton–phonon and electron–phonon interactions\cite{dirnberger2023magneto,mondal2025raman,markina2025interplay,lin2024strong}.
\begin{figure*}[t] 
    \centering
    \includegraphics[width=0.95\textwidth]{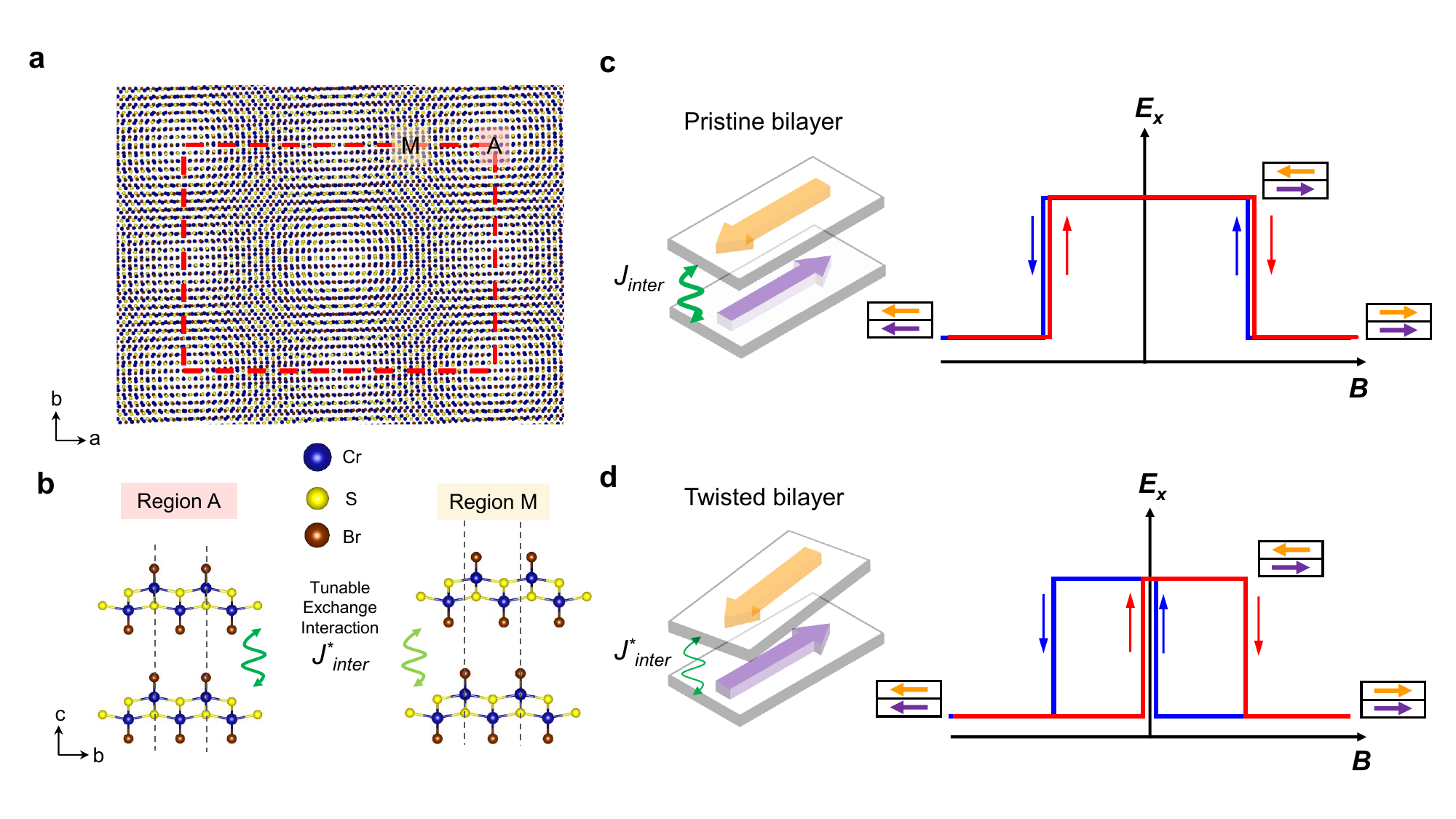}
    \caption{\textbf{Twist engineering and magnetic hysteresis in bilayer CrSBr.}~\textbf{a.} Schematic of the moiré superlattice formed in twisted bilayer CrSBr at a twist angle of \( \theta \approx 3^\circ \), with the moiré unit cell outlined by a red dashed box. ~\textbf{b.} Side view along the \( a \)-axis showing two representative atomic registries—A and M—with distinct stacking configurations. These registries modulate the interlayer exchange interaction.  
    \textbf{c.} Schematic of pristine bilayer CrSBr exhibiting spatially uniform interlayer exchange interaction \( J_{\text{inter}} \). The corresponding exciton energy \( E_x \), that depends on the angle between the two layer magnetisations\cite{wilson2021interlayer}, displays negligible hysteresis under an external magnetic field \( B \) applied along the \( b \)-axis. Red and blue arrows denote the upward and downward field sweep directions, respectively. 
    ~\textbf{d.} In contrast, twisted bilayer CrSBr with spatially averaged interlayer exchange interaction \( J^*_{\text{inter}} \) exhibits pronounced magnetic hysteresis in the \( E_x \)–\( B \) curves. Orange and purple arrows indicate the layer magnetisations of the bilayers. }
    \label{fig:1}
\end{figure*}
\begin{figure*}[t] 
    \centering
    \includegraphics[width=1\textwidth]{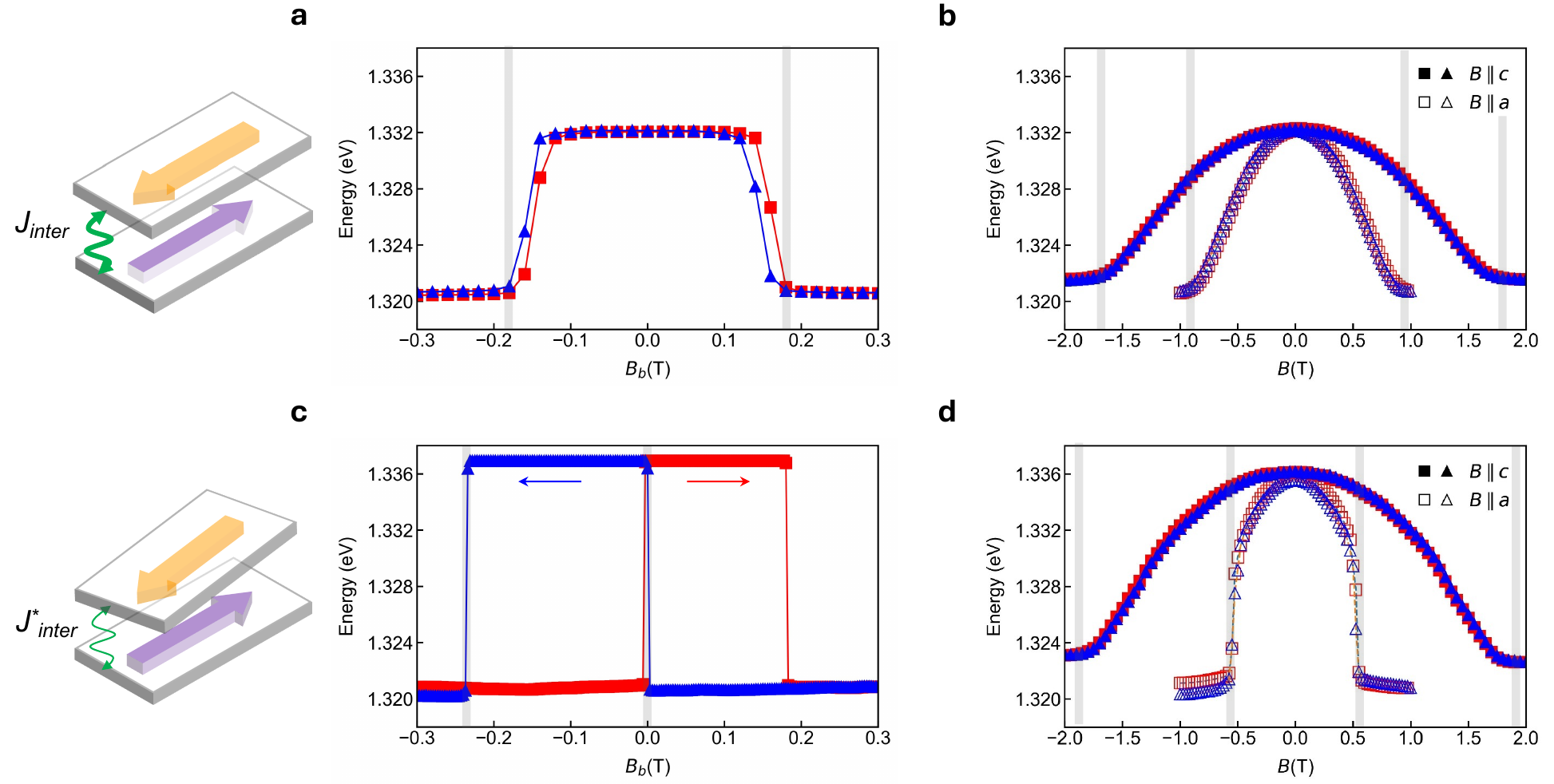}
    \caption{\textbf{Magnetic field dependent exciton energy measured via photoluminescence (PL) spectra in twisted and pristine bilayer CrSBr}.~\textbf{a.}  PL peak energy A-exciton at T=4.7~K of the pristine bilayer under a magnetic field applied along the in-plane $b$-axis. \textbf{b.} same as \textbf{a} but for out-of-plane $c$-axis (full symbols) and $a$-axis (hollow symbols). \textbf{c.}  PL peak energy A-exciton of the twisted bilayer under a magnetic field applied along the in-plane $b$-axis. \textbf{d.} same as \textbf{c} but for out-of-plane $c$-axis (full symbols) and $a$-axis (hollow symbols). In all panels, red squares represent the magnetic field sweep from negative to positive values, while blue triangles indicate the reverse sweep (positive to negative). The measured exciton energy provides direct access to the magnetic configurations (see Fig.~\ref{fig:3}c) due to the former's dependence on the angle between the two layer magnetisations. Switching fields are indicated by grey lines.}
    \label{fig:2}
\end{figure*}

In bilayer CrSBr, the magnetic configuration—and thus the magneto-optical response—is largely determined by the interlayer exchange interaction, which favors antiparallel alignment of the layer magnetisations\cite{wilson2021interlayer,tabataba2024doping}. Here, we demonstrate that a slight angular misalignment between the two layers of a CrSBr bilayer enables control over the interlayer antiferromagnetic exchange (Figs.~\ref{fig:1}a and \ref{fig:1}b). Consequently, a hysteretic and tunable switching between antiparallel and parallel magnetisation states is achieved (Figs.~\ref{fig:1}c and \ref{fig:1}d). Figure~\ref{fig:1}a shows the expected atomic configuration of a twisted bilayer with $\sim$3° twist angle~\cite{liu2025moire}, employed in our experiments. The interlayer exchange is expected to be position dependent following a moir\'e superlattice~\cite{liu2025moire} (Fig.~\ref{fig:1}b). Consequently, the twisted bilayer behaves essentially like its pristine counterpart, but with an effective interlayer exchange $J^*_{\mathrm{inter}}$ obtained by spatial averaging of the position-dependent exchange. Tracking the magnetisation through exciton energy shifts measured by polarization-resolved photoluminescence (PL) provides a highly sensitive probe of magnetic switching\cite{heissenbuttel2025quadratic,serati2023charge,marques2023interplay}. 
\begin{figure*}[t] 
    \centering
    \includegraphics[width=0.95\textwidth]{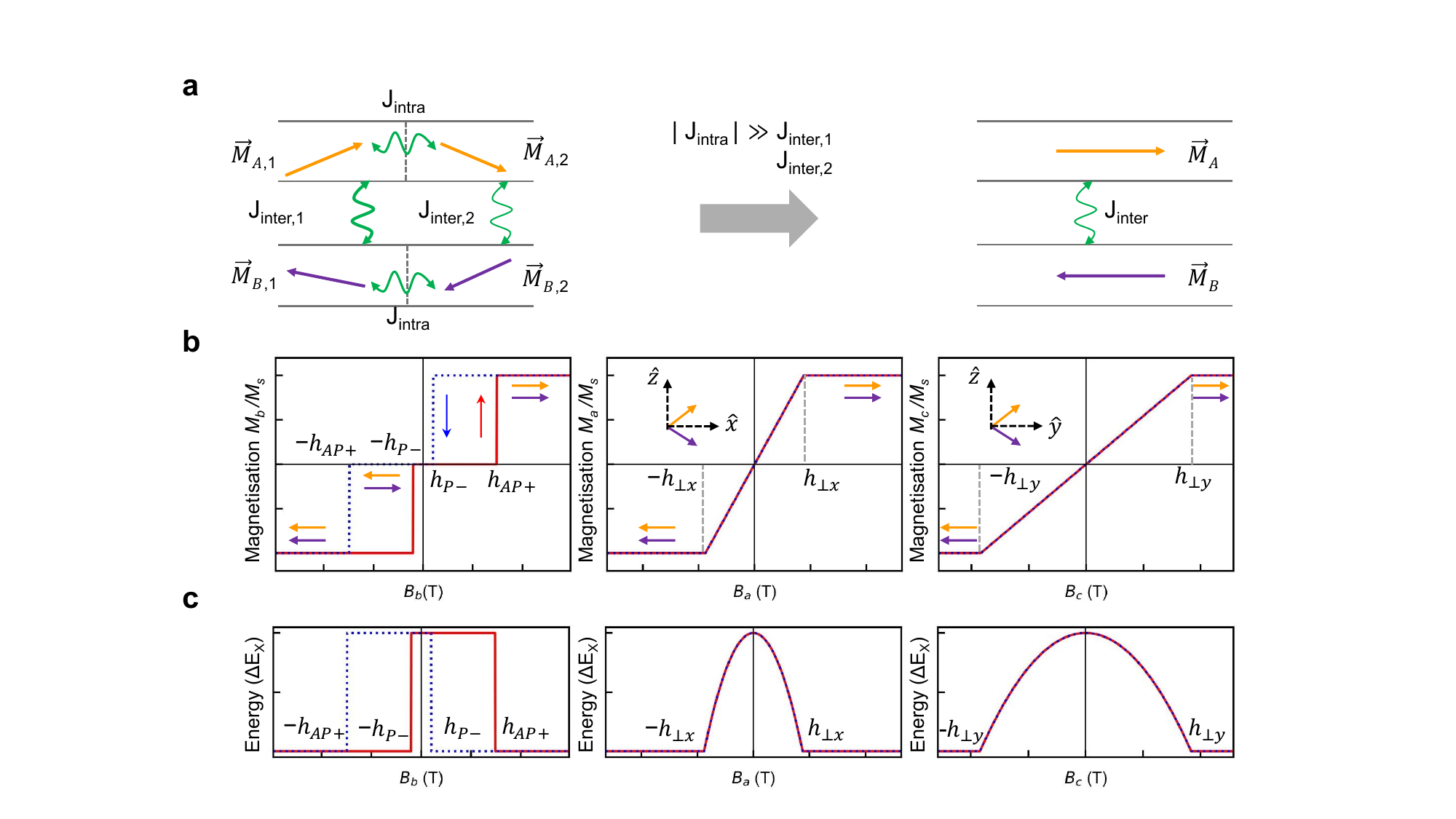}
    \caption{\textbf{Theoretical analysis of magnetic ground state and hysteresis in a twisted bilayer antiferromagnet.} \textbf{a.} Due to the strong intralayer ferromagnetic exchange, the description of position-dependent interlayer exchange (left) arising from a moiré superlattice can be effectively reduced to a simplified two-sublattice model (right) with a spatially averaged effective interlayer exchange. \textbf{b.} Analytically evaluated magnetisation curves \( M/M_s \) as a function of applied magnetic field along the principal crystal axes: magnetic easy axis $b$ (left), intermediate axis $a$ (center), and hard axis $c$ (right) respectively. The red solid and blue dotted curves represent forward and reverse magnetic field sweeps, respectively. Here, hysteresis emerges for magnetic field applied along $b$ because both parallel and antiparallel states are stable in the indicated field range. Switching field values are indicated (see main text) \textbf{c.} The corresponding exciton energy shift $\Delta E_{X}$ evaluated using the relation $\Delta E_{X} \propto \cos^2 \left(\theta/2\right)$ with $\theta$ being the angle between the two sublattice magnetisations~\cite{wilson2021interlayer,heissenbuttel2025quadratic}.} 
    \label{fig:3}
\end{figure*}
While the pristine bilayer is found to manifest no magnetic hysteresis (Fig.~\ref{fig:1}c), the twist-tuned lowering of the interlayer antiferromagnetic exchange makes the parallel magnetisation configuration an energetically stable state for a broader applied field range resulting in magnetic hysteresis (Fig.~\ref{fig:1}d). The good agreement between our experiments and analytic theoretical model thus exposes a new regime for twisted-bilayer magnets manifesting coherence over the moir\'e length scale, reminiscent of similar coherence in correlated states of electronic matter\cite{bistritzer2011moire,cao2018unconventional}. This regime is enabled by the strong intralayer ferromagnetic exchange in CrSBr and the relatively small moir\'e wavelength in our sample. Furthermore, it is complementary to the emergence of spin textures\cite{doi:10.1073/pnas.2000347117} and formation of spatial domains, which also manifests a different hysteresis contributed by the ferromagnetic domains\cite{song2021direct,xu2022coexisting,chen2024twist}. Our demonstrated hysteresis, in sharp contrast, stems from a monodomain switching between antiparallel and parallel magnetisation states that could enable spin filters, memory devices, and reconfigurable logic gates\cite{gong2019two,zhong2020layer} based on antiferromagnetic states.

\indent\textbf{Magnetic hysteresis in twisted versus pristine bilayers}\\
Using low-temperature (4.7 K) photoluminescence (PL) spectroscopy in a confocal microscope\cite{shree2021guide}, we probe the magnetic configurations of pristine and twisted bilayer CrSBr through their excitonic responses under applied magnetic fields. The pristine bilayer exhibits a symmetric, reversible PL response, consistent with previous reports\cite{wilson2021interlayer,tabataba2024doping,boix2024multistep}. Figure~\ref{fig:2}a shows the PL peak energy of the A exciton (see Supplementary Note 3 for spectra) when an in-plane magnetic field is applied along the $b$-axis. Forward and reverse field sweeps, indicated by red squares and blue triangles, are shown for all three directions of the applied magnetic field\cite{krelle2025magnetic}. The A-exciton energy shifts from 1.332 eV to 1.320 eV in both sweep directions, with a switching field of ~0.18 T, resulting in only a very narrow hysteresis profile (Fig.~\ref{fig:2}a).

By contrast, the twisted bilayer shows pronounced hysteresis (Fig.~\ref{fig:2}c). During the forward sweep, the A-exciton energy redshifts from 1.336 eV to 1.320 eV with a switching field of just 3 mT. In the reverse sweep, a comparable redshift occurs at an even lower switching field of ~0 mT, producing a robust hysteresis loop with ferromagnetic order persisting near zero applied field. These stark differences between pristine and twisted bilayers highlight how twist-induced interlayer stacking can stabilize metastable spin states.

To further explore the directional dependence, we applied magnetic fields along the out-of-plane ($c$-axis) and in-plane ($a$-axis) directions (Figs.~\ref{fig:2}b, d). For the pristine bilayer, applying the field along the $a$-axis shifts the A-exciton energy from 1.332 eV (AFM) to 1.320 eV (FM) with a switching field of $\sim$0.9 T in both sweep directions, again without hysteresis. Along the $c$-axis, the exciton shifts from 1.332 eV to 1.318 eV with a switching field of $\sim$1.6 T, also with no hysteresis. In the twisted bilayer, applying the field along the $a$-axis produces a shift from 1.336 eV (AFM) to 1.320 eV (FM) at a switching field of $\sim$0.575 T, with no hysteresis. Along the $c$-axis, the exciton shifts from 1.336 eV to 1.323 eV at $\sim$1.9 T in both directions (see Supplementary Note 4 for details). Thus, hysteresis is observed only in the twisted bilayer for fields along the magnetic easy axis ($b$-axis). Further, the twisted bilayer manifests a magnetic state evolution similar to the pristine sample, but with switching fields altered.

\begin{figure*}[t] 
    \centering
    \includegraphics[width=0.98\textwidth]{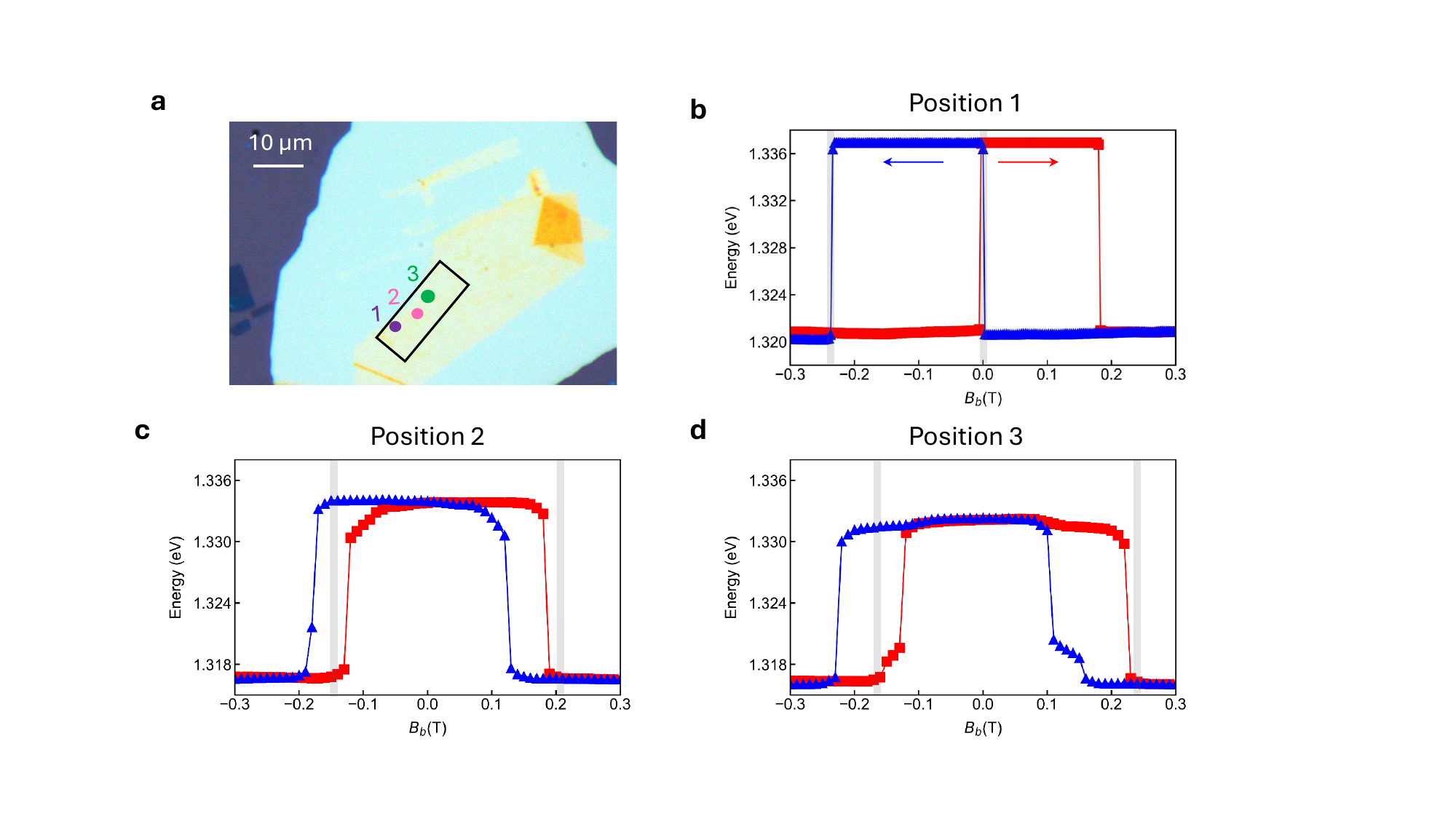}
    \caption{\textbf{Spatially resolved magneto-optical response of a twisted bilayer magnet.} \textbf{a,} Optical microscope image of the fabricated twisted bilayer sample. The black box indicates the region of interest, with three marked positions (1–3) corresponding to the spatial locations where magneto-optical measurements were performed. \textbf{b–d,} Magnetic field-dependent exciton energy measured at positions 1 \textbf{(b)}, 2 \textbf{(c)}, and 3 \textbf{(d)}, under a magnetic field applied along the magnetic easy axis $b$. Red squares represent field sweeps from negative to positive , and blue triangles represent sweeps from positive to negative. The observed hysteresis and energy shifts reflect position-dependent magnetic switching behavior and interlayer exchange coupling in the twisted bilayer. The switching fields $h_{P-}$ and $h_{AP+}$ are indicated by grey lines. While spot 1 manifests only parallel and antiparallel magnetic configurations, spot 2 and spot 3 present evidence for canted state as well, consistent with our theoretical analysis (Supplementary Note 1).}
    \label{fig:4}
\end{figure*}
\indent\textbf{Theory of tunable interlayer exchange dependent magnetic state evolution }\\
The twisting of the bilayer magnet produces a complex position dependent interlayer exchange due to the formation of the moiré superlattice~\cite{liu2025moire} thereby presenting a challenge to theoretical understanding. Fortunately, the intralayer ferromagnetic exchange in CrSBr is much larger than the interlayer exchange and anisotropies. Consequently, for determining the magnetic ground state, the seemingly complex system at hand reduces to a simple two-sublattice model with each layer belonging to one sublattice and the interlayer exchange taking its spatially averaged value [Fig.~\ref{fig:3}a]. This regime relies on large intralayer ferromagnetic exchange and a small moir\'e wavelength, pertinent to our system. Thus, the same model describes the pristine and twisted bilayers, where the twisting enables a tunability of the interlayer exchange. Expressing the magnetic free energy F as~\cite{cham2022anisotropic,dirnberger2023magneto}
\begin{align}
	 \frac{F}{M_s}  = & ~ h_{E} \pmb{m}_{A} \cdot \pmb{m}_{B}   +\frac{h_x}{2}(m_{{A}x}^2+m_{{B}x}^2) \nonumber \\ 
     & ~ +\frac{h_y}{2}(m_{{A}y}^2+m_{{B}y}^2) - \pmb{h}_{\mathrm{ext}} \cdot \left( \pmb{m}_A + \pmb{m}_B  \right),  \label{eq:freegen}
\end{align}
we determine the magnetic state evolution with applied field by evaluating the local minima in the free energy, as detailed in Supplementary Note 1. In Eq.~\eqref{eq:freegen}, $\pmb{m}_{A}$ and $\pmb{m}_{B}$ are the unit vectors along A- and B-sublattice magnetisations, $M_s$ is the saturation magnetisation of each sublattice, and we assume Cartesian x, y, z axes to be aligned with the crystal a, c, and b axes. Further, $h_{E}$, $h_{x}$, $h_y$, and $\pmb{h}_{\mathrm{ext}}$ are $\mu_0$ times the interlayer exchange, intermediate-axis anisotropy, hard-axis anisotropy and externally applied fields.

At zero applied field, the magnet is in its antiparallel state with the sublattice magnetisations aligned along the easy (b or z) axis. For applied magnetic field along the intermediate (a or x) or hard (c or y) axes, the magnetisations slowly cant towards the applied field as its strength is increased until they become parallel at a field of $h_{\perp x} = 2 h_E + h_x$ or $h_{\perp y} = 2 h_E + h_y$, respectively (Fig.~\ref{fig:3}b). For any given field strength for these two axes, a unique magnetic state is allowed. In sharp contrast, for field applied along the easy (b or z) axis, there emerges a parameter space and field range that admits both parallel (P) and antiparallel (AP) magnetic configurations as stable states resulting in hysteresis (see Supplementary Note 1). Consequently, with increasing field strength, AP configuration is maintained until the field reaches a value denoted as $h_{AP+}$ and the magnet switches to the P state. On lowering the field, the magnet switches from P to AP at a lower field of $h_{P-} = 2 h_E - h_x$ resulting in hysteresis. The corresponding magnetisation and exciton energy evolution with applied magnetic field along the three axes is depicted in Fig.~\ref{fig:3}b and c. As detailed in Supplementary Note 1, a canted state also offers a local energy minimum in some range.

If we consider ferromagnetic interlayer exchange in our model,we would only find the P state for any field applied along the easy axes. Our experiments finding switching behavior show that our twisted bilayer sample bears a reduced but clearly antiferromagnetic interlayer exchange. Furthermore, our theoretical analysis provides several switching fields, such as $h_{P-}$ $h_{\perp x}$ and $h_{\perp y}$, which can be read off from the experimental data and employed to calculate the local interlayer exchange as well as anisotropies in the twisted sample, see Supplementary Note 2. 

Based on our parameter extraction (Supplementary Note 2), we conclude that the experimentally observed switching away from the AP configuration at the field of $h_{AP+}$ does not agree with our theoretical analysis. This is the only deviation from our model in a wide range of experiments. We presently do not understand the reasons for this. This issue is also pertinent for the pristine bilayer, including for related materials such as $\mathrm{Cr}\mathrm{I}_3$, for which our theoretical analysis and established parameters predict a hysteresis that is, however, not observed in experiments\cite{wilson2021interlayer,tabataba2024doping,boix2024multistep} (Fig.~\ref{fig:2}a). Thus, besides providing a method for understanding the hysteresis in our twisted bilayers and a reliable extraction of material parameters, our theoretical analysis exposes a key question pertinent to the field of van der Waals magnets.

\indent\textbf{Spatial mapping of interlayer exchange coupling }

To investigate the spatial variation on a length scale much larger than the moir\'e wavelength of the average interlayer exchange coupling in the twisted bilayer, we employ position-resolved PL spectroscopy. Figure~\ref{fig:4} presents hysteresis loops of the A-exciton energy as a function of the applied magnetic field along the $b$-axis at three representative locations across the sample. Each of the examples shown in Fig.~\ref{fig:4}b–d exhibits a distinct hysteresis profile, with notable differences in loop width and switching field. The critical field $h_{P-}$ for forward switching (from negative to positive) varies significantly across positions: $\sim$3~mT at position 1, $\sim$0.14~T at position 2, and $\sim$0.16~T at position 3. The extracted material parameters at each of these positions are presented in Supplementary Note 2.

These spatially dependent features indicate that magnetic switching is influenced by local variations in the interlayer exchange coupling, $J^*_{\mathrm{inter}}$, driven by the stacking configuration. This tunability can be leveraged to engineer domains with tailored magnetic responses and spatially reconfigurable spintronic devices.

Additionally, for each of the three positions, we measured the A-exciton energy as a function of magnetic field applied along the in-plane $a$-axis and the out-of-plane $c$-axis (see Supplementary Note 6). No hysteresis was observed along these directions, confirming that hysteresis is specific to magnetisation along the magnetic easy axis ($b$-axis), consistent with our theoretical analysis based on twist-controlled average interlayer exchange.
Twisting introduces persistent spatial heterogeneity without increasing fabrication complexity therefore providing additional leverage compared to doping, strain or pressure tuning \cite{henriquez2025strain,song2019switching,li2019pressure}.\\

\indent \textit{In summary}, we show that introducing a small twist in bilayer CrSBr enables deterministic control of interlayer exchange, leading to controllable magnetic hysteresis that is absent in pristine bilayers. By combining excitonic spectroscopy with theory, we establish robust design and characterization principles for moir\'e magnets via static magnetic configuration studies. Our findings establish twist engineering as a general route to programmable magnetism in van der Waals materials, with direct implications for the design of reconfigurable spintronic devices, excitonic circuits, and monolayer-scale memory technologies.



\appendix
\section*{Methods}
\indent\textbf{Sample Fabrication}

Bulk CrSBr crystals were fabricated through chemical vapor transport \cite{klein2022control}. The samples were prepared through mechanical exfoliation onto SiO$_2$/Si substrates with an 85 nm SiO$_2$ layer. To probe the topology of the CrSBr flakes Oxford Instruments Cypher atomic force microscope with AC160 cantilevers was used.

\indent\textbf{Atomic Force Microscopy}

Atomic force microscopy measurements were performed at room temperature on a Cypher AFM (Asylum Research/Oxford Instruments, Wiesbaden, Germany). Height images of CrSBr flakes were obtained in AC tapping mode using the cantilever AC160TSA-R3 (300 kHz, 26 N/m, 7 nm tip radius). Images were post-processed with the in-built software features of IGOR 6.38801 (16.05.191, Asylum Research, Santa Barbara, CA, USA).

\indent\textbf{Dry Transfer}

After exfoliating high-quality monolayer CrSBr flakes and hexagonal boron nitride (hBN) flakes onto separate substrates, we employed a dry transfer technique to assemble the twisted heterostructure. The transfer was carried out using a polydimethylsiloxane (PDMS) droplet stamp topped with a thin polypropylene carbonate (PC) film.

The process began with the pickup of the top hBN flake, approximately 30~nm thick. Subsequently, two monolayer CrSBr flakes were sequentially picked up. The pickup of the second CrSBr flake was performed with a rotational misalignment of ~3$^{\circ}$, controlled via a precision rotator mounted on the optical microscope stage. All CrSBr pickups were conducted at a substrate temperature of 100$^{\circ}$C . Finally, a bottom hBN flake (~70~nm thick) was picked up to complete the stack.

The full heterostructure was then released onto a clean Si/SiO$_2$ substrate (with 85~nm oxide thickness). To remove the PC film from the top of the stack, the sample was immersed in chloroform for 5 minutes, followed by rinsing in acetone and isopropanol for 3 and 2 minutes, respectively.

\indent\textbf{Magneto-Optical Spectroscopy}

Optical spectroscopy was performed using a custom-built confocal setup optimized for magneto-optical measurements\cite{shree2021guide}. The sample was mounted in a closed-cycle cryostat (AttoDry 1000XL, attocube systems) equipped with a vector magnet capable of applying fields up to 5~T along the out-of-plane $y$-axis (solenoid) and 2~T along the in-plane $x$- and $z$-axes (split coils). Sample positioning relative to a low-temperature achromatic objective was achieved using piezo-positioners (ANPx101 and ANPz102, attocube systems). Photoluminescence (PL) and differential reflectivity ($DR/R$) measurements were carried out in backscattering geometry at a base temperature of 4.7~K. 
The collected signal was spectrally resolved using a Czerny--Turner spectrograph (SpectraPro HRS-500, Teledyne Princeton Instruments) and detected with a CCD camera (Pylon BRexcelon 100, Teledyne Princeton Instruments). For DR/R measurements, a Tungsten--Halogen lamp (SLS201L/M, Thorlabs) was used as a broadband light source, linearly polarized along the crystal $b$-axis using a nanoparticle-film polarizer and an achromatic half-wave plate. For PL, a HeNe laser polarized along the crystal $a$-axis was used for excitation, with emission collected along the $b$-axis. Excitation powers were varied from 600~nW to 240~$\mu$W. Magnetic field-dependent measurements were performed by first initializing the CrSBr sample in the FM state, followed by sweeping the magnetic field along different crystallographic directions: from $-0.3$~T to $+0.3$~T along the $b$-axis, from $-1$~T to $+1$~T along the $a$-axis, and from $-2$~T to $+2$~T along the $c$-axis for both the twisted and pristine bilayer. Hysteresis loops were obtained by reversing the sweep direction after reaching the maximum field in each case.

\textbf{Data availability}

The data that support the findings of this study are available from the corresponding
authors upon request.

\textbf{Acknowledgments}

K.W. and T.T. acknowledge support from the JSPS KAKENHI (Grant Numbers 21H05233 and 23H02052) , the CREST (JPMJCR24A5), JST and World Premier International Research Center Initiative (WPI), MEXT, Japan. Z.S and K.M. were supported by ERC-CZ program (project LL2101) from Ministry of Education Youth and Sports (MEYS) and by the project Advanced Functional Nanorobots (reg. No. CZ.02.1.01/0.0/0.0/15.003/0000444 financed by the EFRR). S.V. and A.K. thank the German Research Foundation (DFG) for funding via Spin+X TRR 173-268565370, project A13.

\textbf{Author contributions} 
K.M. and Z.S. grew the bulk CrSBr crystals. T.T. and K.W. grew bulk hBN crystals. P.M. and W.L. fabricated few-layer CrSBr samples for optical spectroscopy. P.M. and R.v.K. performed and interpreted AFM measurements. L.K. and R.T. mounted and interfaced the magneto-optical set-up. P.M., W.L., L.K. and B.U. analyzed the optical spectra. S.V and A.K. developed the theoretical model.  B.U. suggested the experiments and supervised the project. P.M., W.L., A.K. and B.U. wrote the manuscript with input from all the authors.

\textbf{Competing interests}: The authors declare no competing interests.

\textbf{Online content}

The data that support the findings of this study are available from the corresponding
authors upon request.

\section*{References}

\clearpage
\onecolumngrid
\appendix

\section*{Supplementary Information}

\newcommand{\akcom}[1]{{\color{magenta} Akash: ``#1''}}
\newcommand{\bucom}[1]{{\color{green} Bernhard: #1}}
\newcommand{\ak}[1]{{\color{black}#1}}

\renewcommand{\thesection}{S\arabic{section}}
\renewcommand{\thesubsection}{S\arabic{section}.\arabic{subsection}}

\renewcommand{\thefigure}{S\arabic{figure}}
\renewcommand{\thetable}{S\arabic{table}}
\renewcommand{\theequation}{S\arabic{equation}}

\setcounter{section}{0}
\setcounter{equation}{0}
\setcounter{page}{1}
\setcounter{figure}{0}

\section{Supplementary Note 1: Theoretical description of the magnetic state and hysteresis}
\label{sec:theory}

In this section, we detail the theoretical framework for describing the magnetic ground state of our twisted CrSBr bilayer as a function of the applied magnetic field along different directions. Our employed two-sublattice model has been successful at describing the pristine bilayer well, while it should be able to describe the magnetization evolution in the twisted bilayer under the assumption that the intralayer ferromagnetic exchange is strong. This assumption is good in CrSBr, but less valid for other van der Waals magnets such as $\mathrm{Cr}\mathrm{I}_3$, which partly explains our qualitatively different results as compared to the previous studies on $\mathrm{Cr}\mathrm{I}_3$~\cite{song2021direct,xu2022coexisting,cheng2023electrically}. Due to the formation of a moir\'e superlattice in the twisted sample, the interlayer exchange becomes position dependent~\cite{liu2025moire}. Understanding the magnetic ground states for this complex position dependent exchange is a formidable theoretical challenge. As shown below, we are able to capture the main experimental observations by introducing an effective interlayer exchange field in our two-sublattice-model. This effective field stems from a local spatial averaging over the relatively small moir\'e unit cell for our twisted bilayer magnet.

In subsection \ref{sec:summary}, we summarize our model, mathematical procedure for the evaluation of magnetic states, and our main results. These findings regarding the magnetic state evolution and hysteresis with applied magnetic field strength are further discussed in subsections \ref{sec:easy}, \ref{sec:intermediate}, and \ref{sec:hard} for the cases of applied field along the easy, intermediate, and hard anisotropy axis, respectively.

\subsection{Summary of the theoretical model and results}
\label{sec:summary}

{\it Free energy description:} We consider a two-sublattice model to describe the antiferromagnetic bilayer. Due to the strong intralayer ferromagnetic exchange in CrSBr~\cite{Scheie2022,Ziebel2024}, the magnetization within each layer is ordered ferromagnetically and represents one sublattice. The interlayer exchange is antiferromagnetic and relatively weak. Thus, it is reasonable to assume that the bilayer twisting by small angles primarily tunes the interlayer antiferromagnetic exchange interaction, although our model allows the anisotropies to vary as well. Further, due to the strong intralayer ferromagnetic exchange, the system is still adequately described as a two-sublattice antiferromagnet with local spatially homogeneous magnetization within each layer.

Within this two-sublattice model, the magnetic free energy $F$ is given by~\cite{dirnberger2023magneto,cham2022anisotropic}:
\begin{align}
	\frac{F}{M_s} & =  \mu_0 H_{E} \pmb{m}_{A} \cdot \pmb{m}_{B}   +\frac{\mu_0 H_x}{2}(m_{{A}x}^2+m_{{B}x}^2) +\frac{\mu_0 H_y}{2}(m_{{A}y}^2+m_{{B}y}^2) - \mu_0 \pmb{H}_{\mathrm{ext}} \cdot \left( \pmb{m}_A + \pmb{m}_B  \right),  \nonumber \\
	 f & =  h_{E} \pmb{m}_{A} \cdot \pmb{m}_{B}   +\frac{h_x}{2}(m_{{A}x}^2+m_{{B}x}^2) +\frac{h_y}{2}(m_{{A}y}^2+m_{{B}y}^2) - \pmb{h}_{\mathrm{ext}} \cdot \left( \pmb{m}_A + \pmb{m}_B  \right).  \label{eq:freegen}
\end{align}
Here, we have chosen the Cartesian coordinate system in such a way that the x, y, and z axes are respectively along the crystal intermediate (a), hard (c), and easy (b) axes. Consequently, $H_{x}$ and $H_{y}$ are positive with $H_{y} > H_{x}$ and represent magnetic anisotropy fields along the x and y axes, respectively. $H_E$ is the interlayer exchange field strength, $\pmb{H}_{ext}$ is the externally applied magnetic field, and $M_s$ is the saturation magnetization of each sublattice. $\pmb{m}_{A,B} \equiv \pmb{M}_{A,B}/M_s$ are the unit vectors along the sublattice A and B magnetizations $\pmb{M}_{A,B}$ assumed to be spatially homogeneous. In the equations above, we have simplified the notation by defining $f \equiv F/M_s$ and $h_{E,x,y,\mathrm{ext}} \equiv \mu_0 H_{E,x,y,\mathrm{ext}}$.

\begin{figure}
	\centering
	\includegraphics[width=80mm]{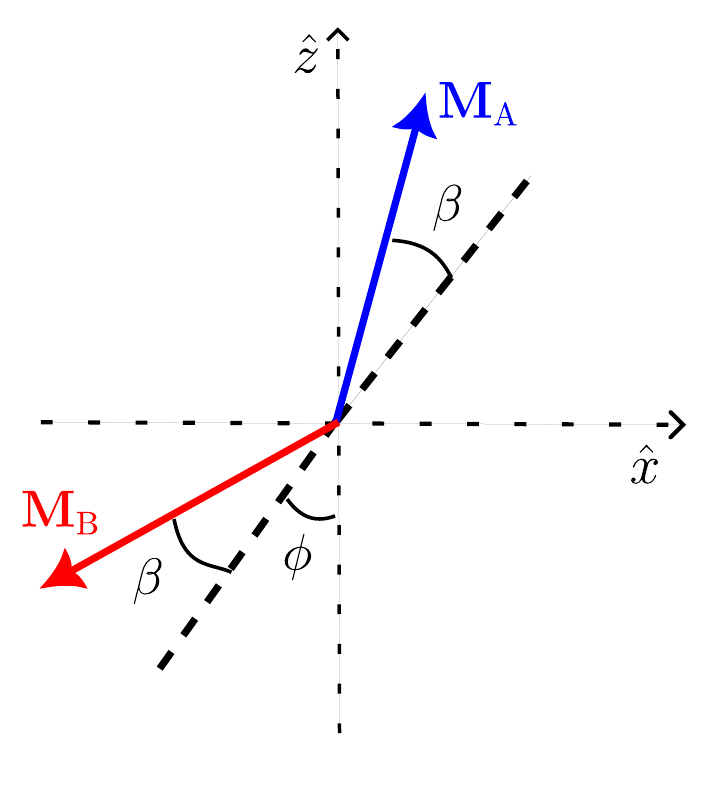}
	\caption{Schematic depiction of the two sublattice magnetizations' configuration. The different values of angles $\beta$ and $\phi$ allow capturing all possible states from antiparallel, parallel, to canted. The schematic depicted adequately captures the cases when the applied field is along the easy ($\hat{z}$) or intermediate ($\hat{x}$) axis. When consider applied field along the hard ($\hat{y}$) axis, the same parametrization of the magnetic configuration works with the x axis replaced by the y axis.}
	\label{fig:model_gen}
\end{figure}

{\it Finding free energy minima:} Any magnetic state that the system attains in equilibrium corresponds to a magnetization configuration that minimizes the free energy expressed via Eq.~\eqref{eq:freegen}. It is possible for the system to have multiple local energy minima, in which case hysteresis emerges due to the applied magnetic field history playing a role in the system configuration~\cite{Gurevich_book,Li2016}. This is the standard approach to understanding the hysteresis in a single-domain ferromagnet with uniaxial anisotropy, for example. Thus, we parametrize the magnetic state of our antiferromagnetic bilayer via angles $\phi$ and $\beta$, as shown in Fig.~\ref{fig:model_gen} and evaluate their values that minimize the free energy Eq.~\eqref{eq:freegen}. A local minimum satisfies the conditions~\cite{Stewart_book}:
\begin{align}
	f_{\phi} = 0, \ f_{\beta} = 0, \ f_{\phi \phi} > 0, \ f_{\beta \beta} > 0, \ \mathrm{and} \ f_{\phi \phi} f_{\beta \beta} - f_{\phi \beta}^2 > 0, 
\end{align}
where $f_{\beta} \equiv \partial f / \partial \beta$, $f_{\phi \phi} \equiv \partial^2 f / \partial \phi^2$ and so on. These conditions provide the equations and inequalities that are solved to obtain the magnetic states that correspond to the local energy minima. Furthermore, we determine the configuration corresponding to the global minimum by comparing the values of the free energy for all the evaluated local minima.

{\it Summary of results:} Following the mathematical procedure described above, we evaluate the values of $\phi$ and $\beta$ that locally minimize the free energy for a given applied magnetic field. At zero applied field, the magnet exists in its expected configuration with antiparallel magnetizations along the z axis.  

For the cases of the applied field along the intermediate (subsection \ref{sec:intermediate}) and the hard (subsection \ref{sec:hard}) axis, we find that with increasing field strength the magnetizations slowly cant towards the applied field direction and become parallel at a large enough field, which is understandably larger for the case of field along the hard axis. However, the system always has a unique configuration that minimizes the energy and thus, there is no hysteresis in these two cases.

When the external field is applied along the easy axis (subsection \ref{sec:easy}), the system remains in the antiparallel (AP) state for low fields. A canted (C) state becomes feasible at somewhat higher fields which evolves into the parallel (P) magnetizations configuration at a large enough field. The unique and remarkable feature, however, is that for a wide range of field values two of the possible three states (AP, C, and P) correspond to local minima in the free energy. This immediately results in a hysteresis since the system stays in the local energy minimum that it occupies due to the history of applied magnetic field and only switches to a different configuration when the occupied local minimum is eliminated at a specific field strength. As the twisting lowers $h_E$, the system enters a parameter space that supports a sizable hysteresis between the AP and P states. Furthermore, our theoretical analysis of the possible magnetic configurations allows a direct extraction of the various material parameters, such as interlayer exchange, characterizing the twisted or pristine bilayer employing the experimentally recorded magnetic configurations as a function of the applied field strength.

\ak{{\it Anomaly and outstanding challenge:} As per our theoretical analysis summarized above and detailed below, hysteresis emerges for the case of magnetic field applied along the easy anisotropy axis. This happens because on increasing the field strength, the system should stay in the AP state as long as it offers a local energy minimum up to an applied field we call $h_{AP+}$ (see Sec.~\ref{sec:easy}). On reducing the field, the system stays in the P state until the field is reduced to $h_{P-}$ (Sec.~\ref{sec:easy}). As per our and another work's material parameter extraction detailed in Sec.~\ref{sec:critfields}, the pristine bilayer should also show this hysteresis because $h_{P-}$ is evaluated to be lower than $h_{AP+}$. However, we do not find any hysteresis in the pristine bilayer. The same is true for studies on pristine $\mathrm{Cr}\mathrm{I}_3$ bilayer~\cite{song2021direct,xu2022coexisting}, which is also adequately captured by our two sublattice model. On a careful look, our theoretical model captures all features of the experimental data very well, but just not the observed field at which the system switches away from the AP state. 

In summary, our theoretical two-sublattice model predicts hysteresis for both pristine and twisted bilayers. It seems to describe our twisted samples better than the pristine ones, even though the latter have been successfully described using this model for many phenomena. We obtain very good consistence between our theory and experiments if we assume that the experimentally observed field (called $h_{AP+}$ in our theoretical analysis) for switching away from the AP state is somehow not adequately captured by the theory. The reasons for this intriguing anomaly are presently not understood.

Despite this intriguing anomaly, that we leave for future research, our theory and experiment agree very well and allow for an effective extraction of the material parameters for the twisted magnets using the static magnetization data.}

\subsection{Magnetic field applied along the easy axes}
\label{sec:easy}

\begin{figure}
	\centering
	\includegraphics[width=100mm]{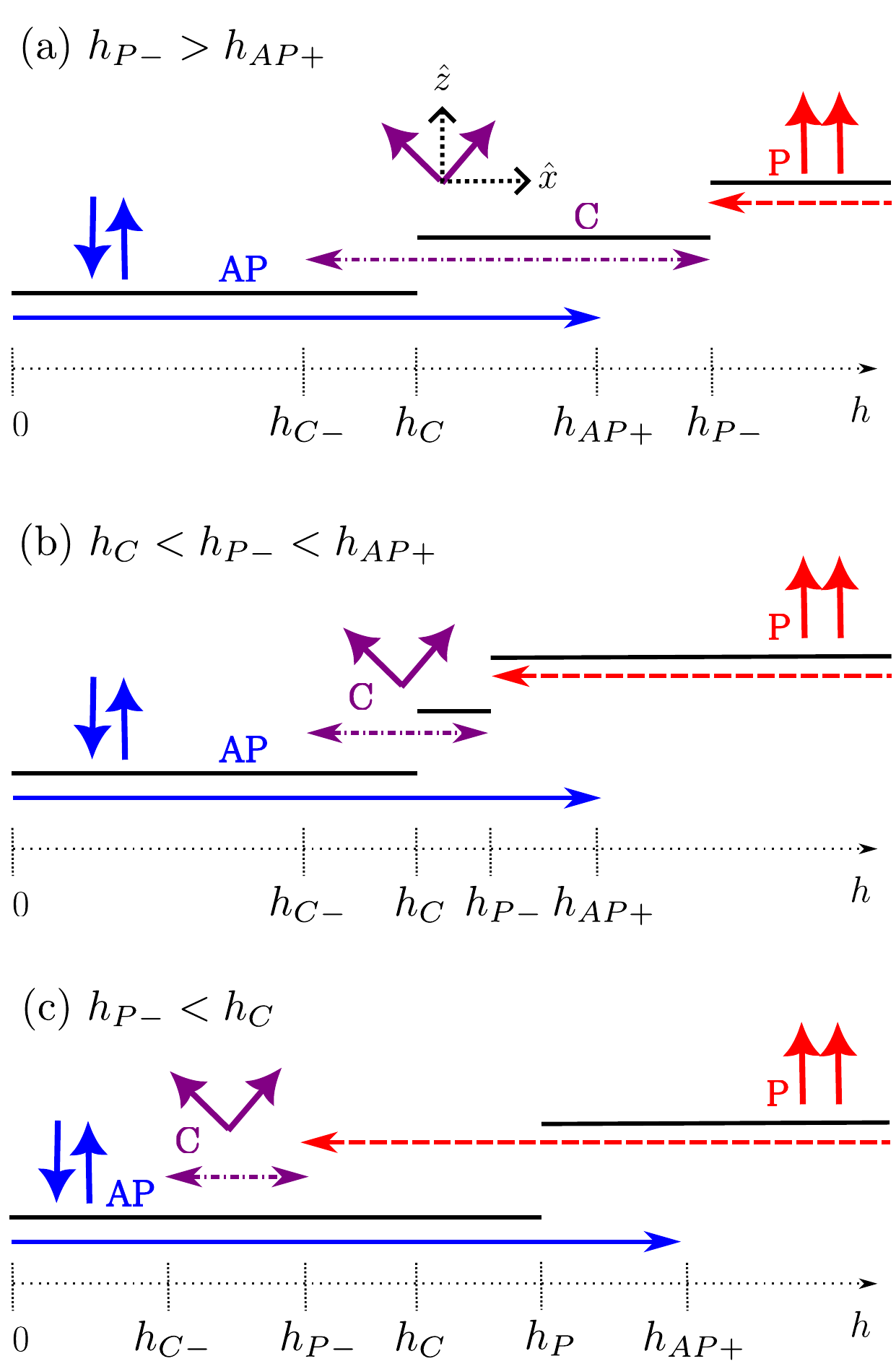}
	\caption{Schematic depiction of allowed magnetic configurations for a given strength of field $h$ applied along the easy (z) axis. The various characteristic field values are defined in Table \ref{tab:easy}. Panels (a), (b) and (c) show the situations for three different hierarchies between the characteristic field values. The blue solid, magenta dashed dotted, and red dashed lines depict the presence of antiparallel (AP), canted (C), and parallel (P) magnetic configurations, respectively, as stable allowed states corresponding to local minima in the free energy. The black solid line above highlights the state corresponding to the global minimum in energy for any given applied field $h$. Magnetic hysteresis emerges when there are multiple stable states for any range of applied field values. \ak{Based on the material parameters (Sec.~\ref{sec:critfields}), panel (c) is relevant to our CrSBr samples.}}
	\label{fig:easy_states}
\end{figure}

Let us consider an applied magnetic field such that $\pmb{h}_{\mathrm{ext}} = h \hat{z}$. Furthermore, assuming the configuration depicted in Fig.~\ref{fig:model_gen}, the magnetization unit vectors become 
\begin{align}
	\pmb{m}_{A} & = \cos{(\phi-\beta)}\hat{z} + \sin(\phi-\beta) \hat{x}, \nonumber \\
	\pmb{m}_{B} & = -\cos{(\phi+\beta)}\hat{z} - \sin(\phi+\beta) \hat{x}. \label{eq:mconf}
\end{align}
Substituting these in Eq.~\eqref{eq:freegen}, we obtain the free energy expression $f_z$ relevant for magnetic field applied along the easy axis:
\begin{align}\label{eq:fz}
	 f_z & = - h_{E} \cos(2{\beta}) + \frac{h_x}{2} - \frac{h_x}{2} \cos{(2\phi)} \cos{(2\beta)} - 2h\sin{\phi}\sin\beta.
\end{align}
The possible magnetic states are obtained via the values of $\phi$ and $\beta$ which correspond to the local minima in the free energy expression Eq.~\eqref{eq:fz} above. This determination of local and global minima in energy is carried out by following the standard procedure and solving the equations outlined in Sec.~\eqref{sec:summary}. We directly discuss the results, which are summarized in Fig.~\ref{fig:easy_states}.   

\begin{table}
\begin{center}
	\begin{tabular}{ |c|c|c|c|c|c| } 
		\hline
		Characteristic field & $h_{C-}$ & $h_{AP+}$ & $h_{P-}$ & $h_{C}$ & $h_{P}$  \\ 
		\hline
		 Value in terms of material properties & $h_{P-} \sqrt{\frac{h_x}{2 h_E + h_x}}$ & $\sqrt{h_x (2 h_E + h_x)}$ & $2 h_E - h_x$ & $\sqrt{h_x h_{P-}}$ & $h_E$ \\  
		\hline
	\end{tabular}
\caption{For the case of applied magnetic field along the easy axis, the various characteristic fields that separate the applied field axis into regions with different magnetic states (see Fig.~\ref{fig:easy_states}) are related to the material properties characterizing the magnetic free energy Eq.~\eqref{eq:freegen}.}
\label{tab:easy}
\end{center}
\end{table}

Let us first consider the case when interlayer antiferromagnetic exchange, represented by $h_{E}$, is larger than a specific value that we discuss below [Fig.~\ref{fig:easy_states}(a)]. In this case, the system evolves from being in the antiparallel (AP) state ($\phi = 0, \beta = 0$) to the canted (C) state ($\phi = \pi/2, 0 < \beta < \pi/2$) to the parallel (P) state ($\phi = \pi/2, \beta = \pi/2$), with increasing applied magnetic field $h$. The AP state presents a local minimum in the free energy for $h < h_{AP+}$ (see Table \ref{tab:easy}) and the magnet is thus allowed to be in AP state for this range of applied magnetic field. For $ h_{C-} < h < h_{P-}$, the C state presents a local minimum and is allowed. Finally, for $h > h_{P-}$, the system is in P state. Interestingly, in the field range $h_{C-} < h < h_{AP+}$, both AP and C present local minima in the free energy. Due to the multiple allowed states, the system is expected to manifest hysteresis in this field range. The situation is summarized in Fig.~\ref{fig:easy_states}(a), where the global energy minimum is also represented. This case is relevant when $h_{P-} > h_{AP+}$, \ak{which is the case for conventional antiferromagnets with the antiferromagnetic exchange being much larger than the anisotropies}. The canted state is described by $\phi = \pi/2$ and
\begin{align}
	\sin \beta = \frac{h}{h_{P-}}.
\end{align}
Here, $h_{P-}$ and other characteristic field values discussed above have been defined in Table \ref{tab:easy}.

The magnetic state evolution is schematically depicted in Fig.~\ref{fig:easy_states}(b) for the case $h_C < h_{P-} < h_{AP+}$. Here, $h_C$ is the field at which the canted state becomes the global energy minimum with an increasing applied field value. When $h_{P-}$ is further reduced, for example, due to a decrease in the interlayer exchange $h_{E}$ on account of bilayer twisting, we obtain the state evolution as shown in Fig.~\ref{fig:easy_states}(c). Now, a new characteristic field $h_{P}$ emerges. At this value, the P state starts being the global energy minimum with increasing applied field. 

The evolution described by Fig.~\ref{fig:easy_states}(a) is typical for crystalline antiferromagnets (AFMs) which have $h_{E} \gg h_{x}$. For our experiments with CrSBr, the situation depicted in Fig.~\ref{fig:easy_states} (c) is pertinent, \ak{as confirmed by our explicit evaluation of the characteristic fields (Sec.~\ref{sec:critfields}) based on the extracted material parameters for pristine as well as twisted bilayers.}

Let us consider the hysteretic state evolution observed in our experiments considering Fig.~\ref{fig:easy_states} (c). At zero field, we have AP configuration. As the field is increased, the magnet should remain in the AP state until $h = h_{AP+}$ when the AP configuration stops being a local minimum in energy and only P configuration presents a stable state. The magnet remains in P on further increase of the field. Considering the backwards trajectory of lowering the applied magnetic field, the system remains in the P state until the field reaches $h = h_{P-}$ at which P stops presenting a local energy minimum. At this stage, the system can switch to AP or C configurations. Our experiments indicate that it switches to the AP configuration \ak{in most cases}. This can be rationalized by considering that in going from P to AP configuration, the magnet has to reverse one of the two sublattice magnetizations, which is easier to happen stochastically. For the system to achieve the C state, both sublattice magnetizations would need to be reoriented in a somewhat coordinated fashion, which seems harder to be accomplished by the thermal stochastic forces. Our theoretical model is unable to fully capture this stochastic switching, which could be non-deterministic. \ak{The state (AP or C) to which the system switches going away from P on lowering the applied field is not deterministic, as discussed above, and we find slightly different behaviors in our data recorded at spots 1, 2, and 3 (see Fig.~4 in the main text). Nevertheless, all these behaviors are anticipated and can be understood within our theoretical model.}

\ak{As discussed above, our theoretical model expects a certain value (see Table \ref{tab:easy}) for the field $h_{AP+}$ at which the system switches away from the AP state on increasing the applied field. However, the experiment finds switching at a different value, which is the only inconsistency between our theory and experiment. In other words, our experimentally measured value of $h_{AP+}$ is not consistent with its theoretically evaluated value based on the material parameters (Sec.~\ref{sec:critfields}). This anomaly remains even for the case of pristine bilayer where the measured value of $h_{AP+}$ coincides with $h_{P-}$ thereby resulting in no hysteresis. At present, we do not understand the reason for this anomaly. Nevertheless, by acknowledging this unreliability of the measured $h_{AP+}$ and thus disregarding this value, we are still able to extract all material parameters for our pristine and twisted bilayers by comparing theory and experiment, as detailed in Sec.~\ref{sec:critfields}.}

\subsection{Magnetic field applied along the intermediate axis}
\label{sec:intermediate}

\begin{figure}
	\centering
	\includegraphics[width=100mm]{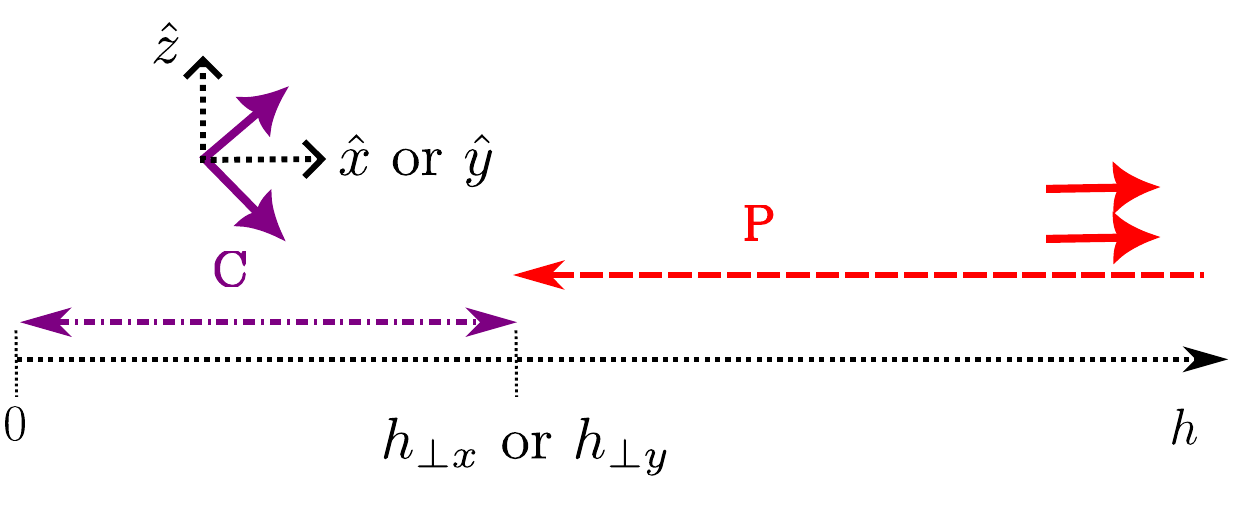}
	\caption{Schematic depiction of allowed magnetic configurations for a given strength $h$ of the field applied along the intermediate (x) or hard (y) axis. A unique state is allowed for any given field strength. The system is in a canted (C) state until $h$ reaches $h_{\perp x} \equiv 2 h_E + h_x$ ($h_{\perp y} \equiv 2 h_E + h_y$) for the case of field along intermediate (hard) axis. For larger fields, the system is in parallel (P) state.}
	\label{fig:inter}
\end{figure}

Considering the applied magnetic field along the intermediate anisotropy (x) axis, we can repeat the analysis carried out in the previous subsection. The free energy function now takes the form:
\begin{align}\label{eq:fx}
	f_x & = - h_{E} \cos(2{\beta})  + \frac{h_x}{2} - \frac{h_x}{2}\cos{(2\phi)}\cos{(2\beta)} + 2h\cos{\phi}\sin\beta,
\end{align}
in terms of $\phi$ and $\beta$ (Fig.~\ref{fig:model_gen}). Following the procedure described in Sec.~\ref{sec:summary}, we minimize the free energy above with respect to $\phi$ and $\beta$ thereby obtaining the magnetic states for any given applied field strength $h$. In this case, we find a unique magnetic configuration for any field value and thus, no hysteresis is expected. The system manifests a canted (C) state with $\phi = \pi$ and
\begin{align}
	\sin \beta & = \frac{h}{h_{\perp x}},
\end{align} 
for $0 < h < h_{\perp x}$ and the parallel (P) state with $\phi = \pi$ and $\beta = \pi/2$ for $h \geq h_{\perp x}$. Here, $h_{\perp x} = 2 h_E + h_x$, and the magnetic state evolution with applied field is schematically summarized in Fig.~\ref{fig:inter}.

\subsection{Magnetic field applied along the hard axis}
\label{sec:hard}

When the magnetic field is applied along the hard axis, the sublattice magnetizations are no longer pointing in the x-z plane, as was the case in the two cases above. Instead, the applied magnetic field forces the sublattice magnetization to orient in the y-z plane. Nevertheless, we may employ the same parametrization of the general magnetic state as schematically depicted in Fig.~\ref{fig:model_gen} with one small change: x axis is replaced by y axis. With this new parametrization, we may follow the same mathematical procedure for obtaining and minimizing the free energy as discussed in the previous section. The obtained result is also nearly identical to the case of magnetic field applied along the intermediate axis (Sec.~\ref{sec:intermediate}) with one key difference: $h_{\perp x}$ is now replaced by $h_{\perp y} \equiv 2 h_E + h_y$. The magnetic state evolution is conveniently summarized in Fig.~\ref{fig:inter} together with the case of field along the intermediate axis.

\clearpage
\section{Supplementary Note 2: Extracting tuned exchange and anisotropies from experimental data}
\label{sec:critfields}

\ak{Here we provide the procedure and a table to give the field values characterizing our CrSBr samples extracted from experiments. 

{\it Measured switching fields and error analysis:}} The measured switching fields $h_{P-}$, $h_{AP+}$, $h_{\perp x}$ and $h_{\perp y}$ for pristine and twisted bilayer (Spot 1) are \ak{marked} in the Figure 2 of the main text. The switching fields of twisted bilayer Spot 2 and 3 are shown in Figure 4 of the main text and in Figure \ref{fig:figS8} and \ref{fig:figS9}. The measurements were performed with a magnetic field stepsize of 10 mT along the $b$-axis. But for extracting the critical fields, in particular for Spot 2 and 3 of the twisted sample, the error is larger, as the change in magnetisation does not occur as a step-like function. For the $a$ and $c$ axis the error bars can be set equal to the magnetic field step-size. We therfore estimate the error of extracted fields to be $\pm$15 mT, $\pm$25 mT, $\pm$50 mT under the external magnetic fields parallel to the $b$, $a$, and $c$ axis, respectively. 

\ak{
{\it Extracting material properties:} We have four measured fields - $h_{P-}$, $h_{AP+}$, $h_{\perp x}$ and $h_{\perp y}$, and only three variables to be determined - $h_{E}$, $h_x$, and $h_y$. Thus, the system of equations is overdetermined. On careful analysis, we find that employing the following expressions
\begin{align}
h_{P-} & = 2 h_E - h_x, \\
h_{\perp x} & = 2 h_E + h_x, \\
h_{\perp y} & = 2 h_E + h_y,
\end{align}
and solving for the three desired field values ($h_{E}$, $h_x$, and $h_y$) yields a good procedure to extract these. We find that employing the theoretical expression for $h_{AP+}$ (Table \ref{tab:easy}) in the extraction procedure produces unphysical and anomalous field values, which do not compare at all with other determinations in the literature. This, and several other indications, convince us that the experimentally recorded value of $h_{AP+}$ does not correspond to its theoretical evaluation within our model for reasons that we do not understand at present. Furthermore, in order to validate our procedure for extracting the field values characterizing the material, we compare our results with the previous results on bulk CrSBr~\cite{cham2022anisotropic}. These are also reported and discussed in the table below and we find a good agreement. We further note that the use of ferromagnetic resonance frequencies for fitting and extracting the field values in Ref.~\cite{cham2022anisotropic} produces some deviation from our extracted values. If we use the static measurements from Ref.~\cite{cham2022anisotropic} to extract the fields characterizing the material, we get a better agreement with our extracted values. At the same time, this shows that fitting the ferromagnetic resonance frequencies data and static magnetization data using the same free energy produces slightly different extracted values of the fields characterizing the material. 
}



\begin{table}[htbp]
    \centering
\caption{Extraction of \ak{fields (in Tesla) characterizing the material} from experimental data. }
\label{tab:my_label}
\begin{threeparttable}
    \begin{tabular}{|l|l|l|l|l|l|l|l|l|l|l|}\hline
 & \multicolumn{4}{|c|}{Measured}& \multicolumn{3}{|c|}{Extracted} & \multicolumn{3}{|c|}{Evaluated}\\\hline 
         Sample&  $h_{P-}$&   $h_{AP+}$&$h_{\perp x}$&  $h_{\perp y}$&  $h_x$&  $h_y$&  $h_E$& $h_{AP+}$\tnote{i}~~& $h_{C-}$\tnote{ii}~~&$h_C$\tnote{iii}\\ \hline  
         Twisted (Spot 1)&  0&   0.237&0.575&  1.900&  0.288&  1.612&  0.144 & 0.407& 0&0\\ \hline  
         Twisted (Spot 2)&  0.140&   0.210&0.500&  1.800&  0.180&  1.480&  0.160& 0.300& 0.084&0.159\\ \hline  
         Twisted (Spot 3)&  0.160&   0.250&0.600&  1.800&  0.220&  1.420& 0.190& 0.363& 0.097&0.188\\ \hline
 Pristine bilayer& 0.180& 0.180& 0.900& 1.600& 0.360& 1.060&0.270 & 0.569& 0.114&0.255\\ \hline\hline
 Bulk (Ref.~\cite{cham2022anisotropic}) \tnote{iv}~~~~& \multicolumn{4}{|c|}{N/A}& 0.383 & 1.30 & 0.395 & 0.670&0.233&0.395\\\hline
    \end{tabular}
    \begin{tablenotes}
      \footnotesize 
      \item[i] $h_{AP+}=\sqrt{h_x(2h_E+h_x)}$ \ak{(Table \ref{tab:easy})}
      \item[ii] $h_{C-}=h_{P-}\sqrt{\frac{h_x}{2h_E+h_x}}$ \ak{(Table \ref{tab:easy})}
      \item[iii] $h_C=\sqrt{h_xh_{P-}}$ \ak{(Table \ref{tab:easy})}
      \item[iv] \ak{The extracted values of the fields here have been taken directly from Ref.~\cite{cham2022anisotropic}. These field values were extracted by fitting the ferromagnetic resonance data at 5K to the corresponding theoretical calculation based on the same free energy that we have employed [Eq.~\ref{eq:freegen}] in our analysis of the magnetic ground state and hysteresis.}
    \end{tablenotes}
\end{threeparttable}
\end{table}

\clearpage
\section{Supplementary Note 3: Pristine Bilayer}
\label{sec:prist}



\begin{figure*}[htbp]
\includegraphics[width=1\linewidth]{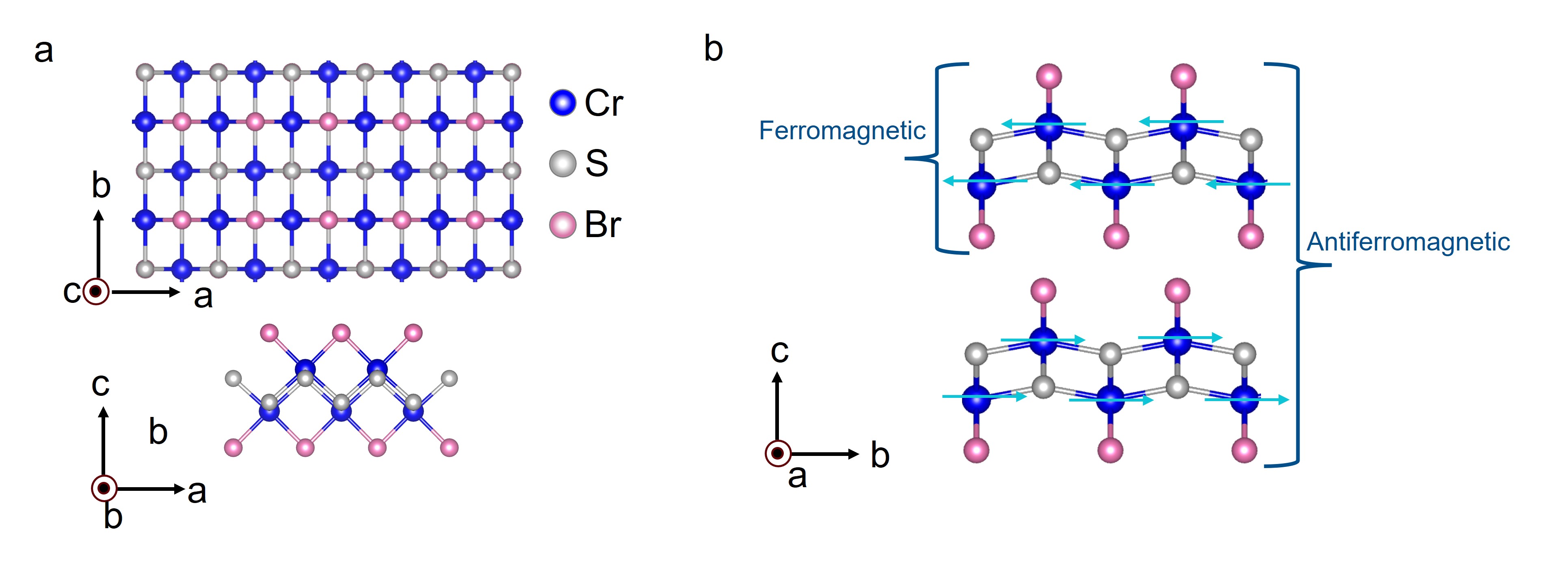}
\caption{\label{fig:figS1} \textbf{Crystal structure of CrSBr.}
(a) Crystal structure oriented along the c- and b-axes, showing the layered arrangement of Cr, S, and Br atoms.
(b) Projection along the a-axis, highlighting the in-plane magnetic spin alignment. Within a single layer, the Cr atoms exhibit ferromagnetic ordering, while adjacent layers are antiferromagnetically coupled. This layered magnetic structure underlies the anisotropic magnetic and magneto-optical properties of CrSBr.}
\end{figure*}

\begin{figure*}[htbp]
\includegraphics[width=0.65\linewidth]{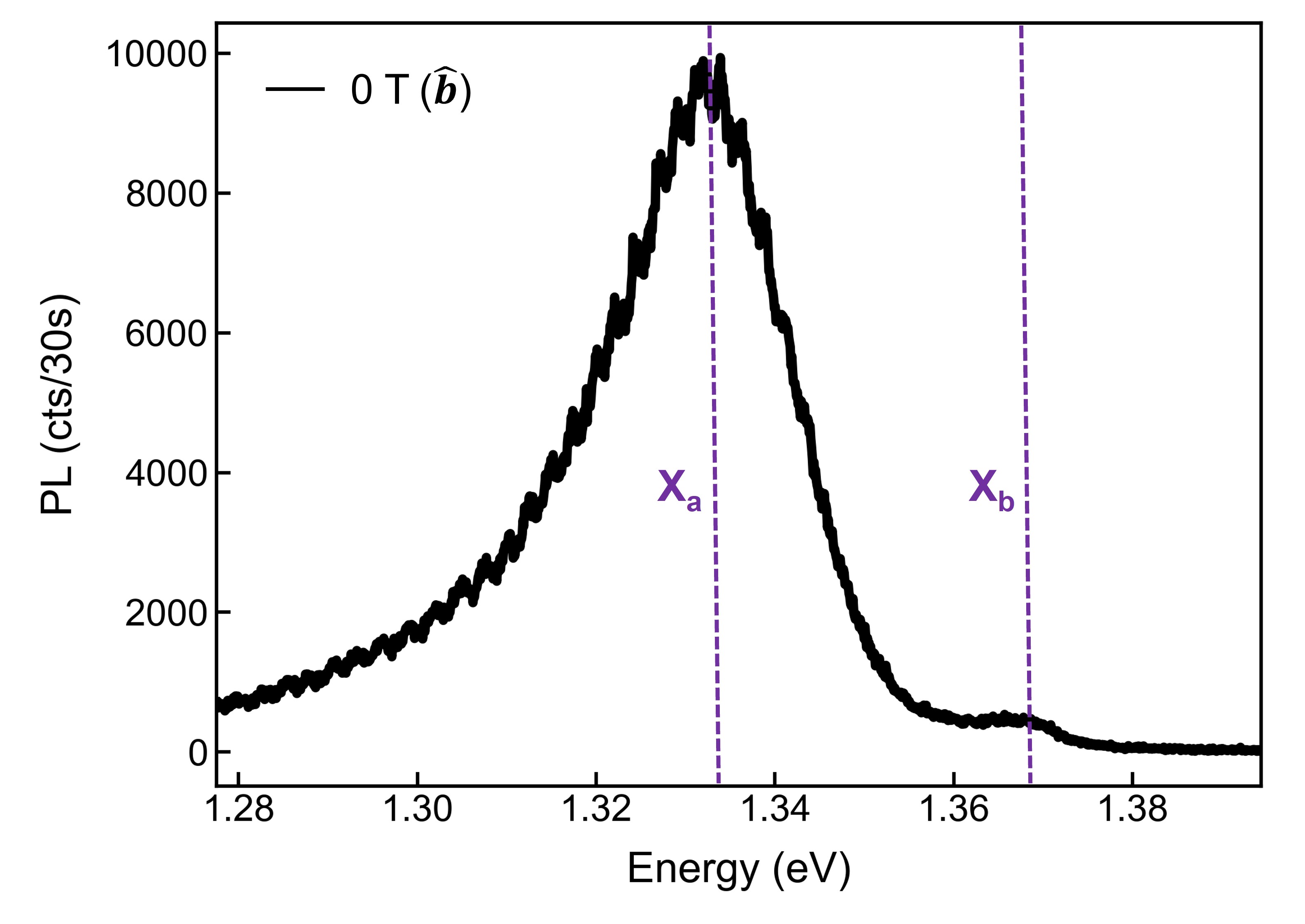}
\caption{\label{fig:figS2} \textbf{Photoluminscence spectrum of pristine bilayer CrSBr.}
Two excitonic species, the A and B excitons, are observed in the spectra of pristine bilayer CrSBr at 0~T and 4.7~K, appearing at energies of 1.332~eV and 1.367~eV, respectively.}
\end{figure*}

\begin{figure*}[htbp]
\includegraphics[width=1\linewidth]{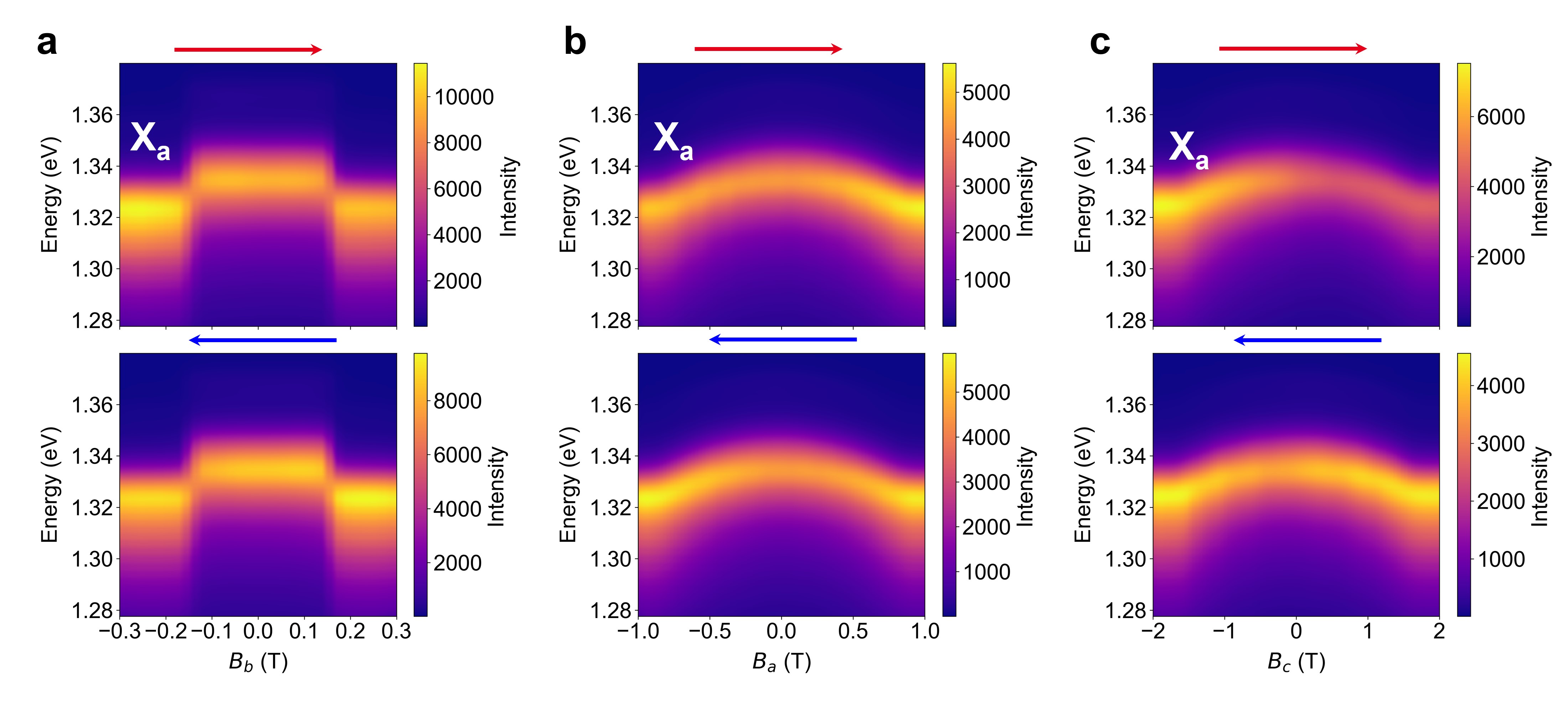}
\caption{\label{fig:figS3} \textbf{Magneto-optical measurements on the pristine bilayer under magnetic fields applied along specific crystallographic directions.}
Panels \textbf{a}, \textbf{b}, and \textbf{c} show the magnetic field dependence along the $b$-, $a$-, and $c$-axes, respectively. For each orientation, the top row displays the data for increasing field (from negative to positive), and the bottom row for decreasing field (from positive to negative). The A-exciton becomes brighter as the system transitions from the antiferromagnetic to ferromagnetic regime. }
\end{figure*}

Supplementary Note \ref{sec:prist} illustrate thes structural, optical, and magneto-optical properties of  pristine bilayer CrSBr, as we compare in the main text the pristine with a twisted bilayer (see Supplementary Note \ref{sec:XinTwist}). The crystal structure along multiple axes, highlighting the layered arrangement of Cr, S, and Br atoms and the anisotropic magnetic ordering within and between layers is shown in Fig.~\ref{fig:figS1}.  The photoluminescence (PL) spectrum at 4.7~K and zero magnetic field of pristine bilayer CrSBr in Fig.~\ref{fig:figS2} shows two exciton species. The peaks located at 1.332 eV and 1.37eV are assigned to the A and B exciton, respectively, consistent with previous reports\cite{wilson2021interlayer,tabataba2024doping}. Detailed magneto-optical measurements under applied magnetic fields along the three crystallographic axes demonstrate the evolution of the A-exciton intensity as the system transitions between antiferromagnetic and ferromagnetic regimes in  Fig.~\ref{fig:figS3}, for details of the vector magnet set-up see \cite{krelle2025magnetic}. We do not observe hysteresis of the magnetization for magnetic field sweeps for the pristine bilayer. 


\clearpage
\section{Supplementary Note 4: Excitonic transitions in Twisted Bilayer CrSBr.}
\label{sec:XinTwist}

\begin{figure*}[htbp]
\includegraphics[width=0.6\linewidth]{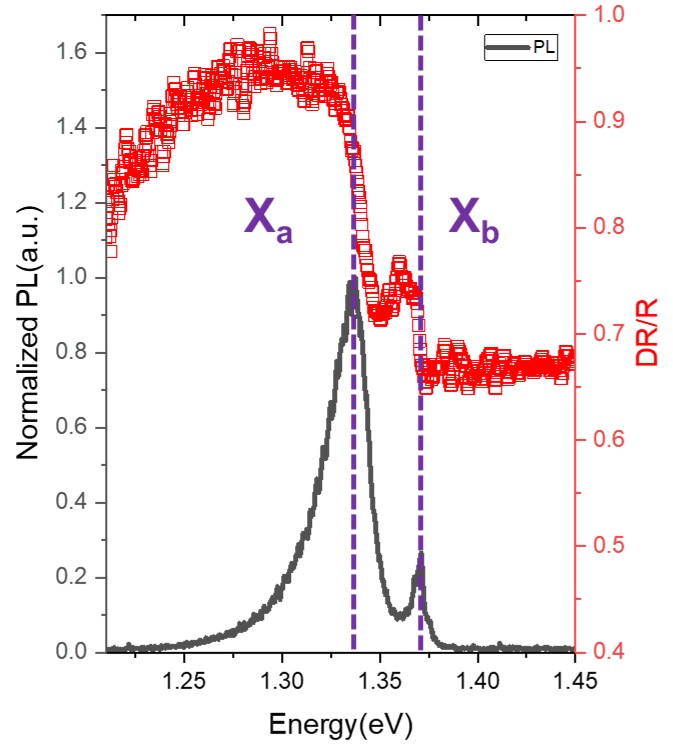}
\caption{\label{fig:figS4} \textbf{Twisted bilayer photoluminescence and differential reflectance(DR/R) measurements at 4.7~K at zero applied magnetic field.} The black curve represents the photoluminescence (PL) spectrum of the twisted bilayer, showing two distinct peaks corresponding to the A-exciton at 1.336~eV and the B-exciton at 1.370~eV. The red curve shows DR/R measured at the same sample position.}
\end{figure*}

In the twisted bilayer at B$=0$~T and T$=4.7$~K, we newly observe distinct peaks in PL at 1.336 eV and 1.37 eV. Both transitions are also clearly visible in differential reflectance spectra $DR/R$ (Fig.~\ref{fig:figS4}), indicating their high oscillator strength\cite{shree2021guide}. Both the PL and reflectivity spectra exhibit two clear excitonic features, indicating consistent optical responses from the twisted bilayer structure.

\begin{figure*}[htbp]
\includegraphics[width=1\linewidth]{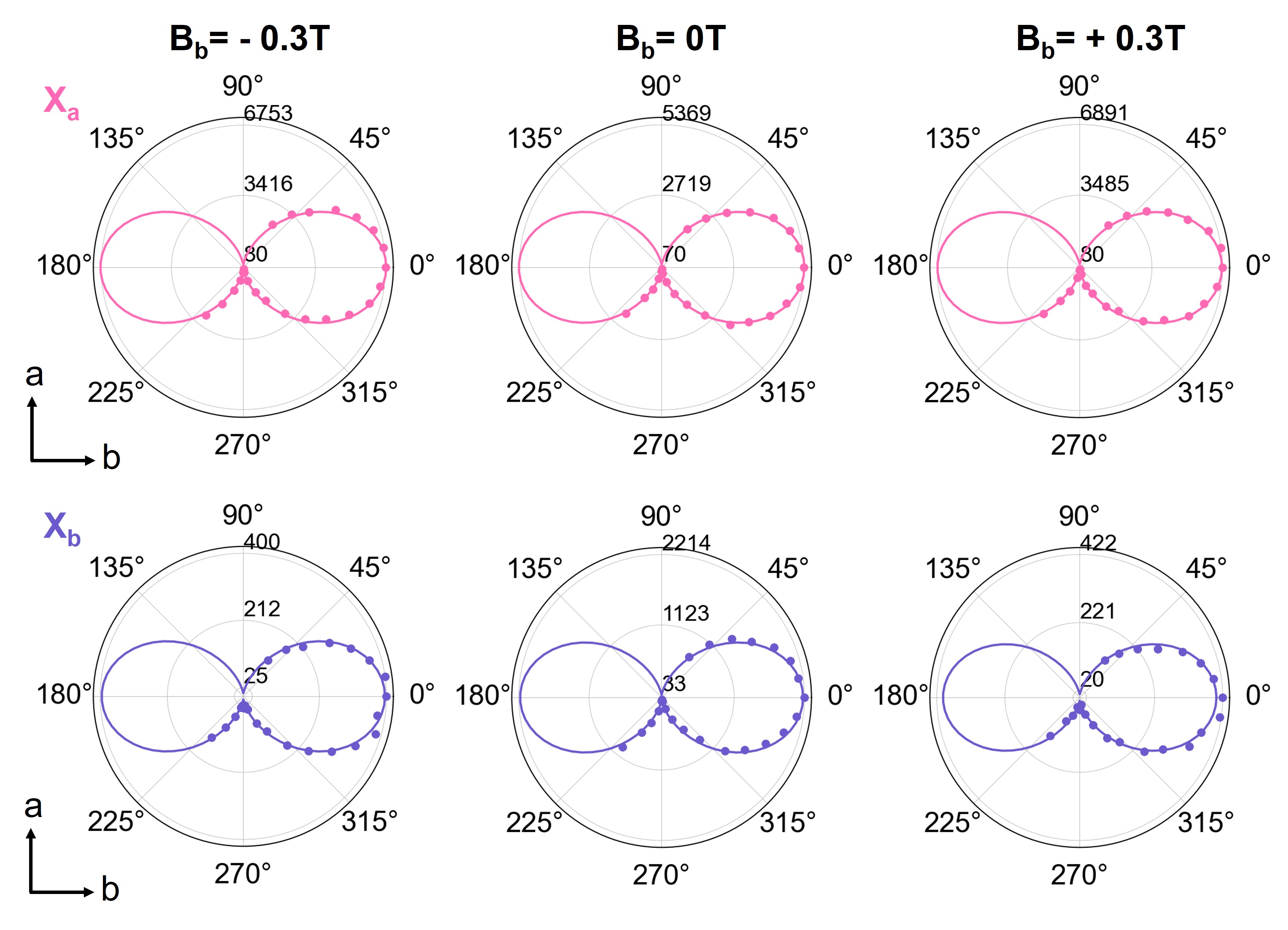}
\caption{\label{fig:figS5} \textbf{Polarization-resolved photoluminescence (PL) from the twisted bilayer sample at T=4.7~K using a half-wave plate (HWP) to change the excitation polarization direction.} Top panel shows the polarization dependence of A-exciton and bottom panel of the B-Exciton. The left panel shows the polarization-resolved PL at a magnetic field of –0.3~T along the b-direction, the middle panel corresponds to zero field, and the right panel shows the measurement at +0.3~T along the b-direction. Pink dots represent the A-exciton, while purple dots correspond to the B-exciton.}
\end{figure*}

We examined the polarization properties of these excitons in both the ferromagnetic (±0.3~T along $b$) and antiferromagnetic (0~T) states. Regardless of the spin configuration, both excitons remain strongly polarized along the $b$ crystallographic axis (Fig.~\ref{fig:figS5}).

\begin{figure*}[htbp]
\includegraphics[width=1\linewidth]{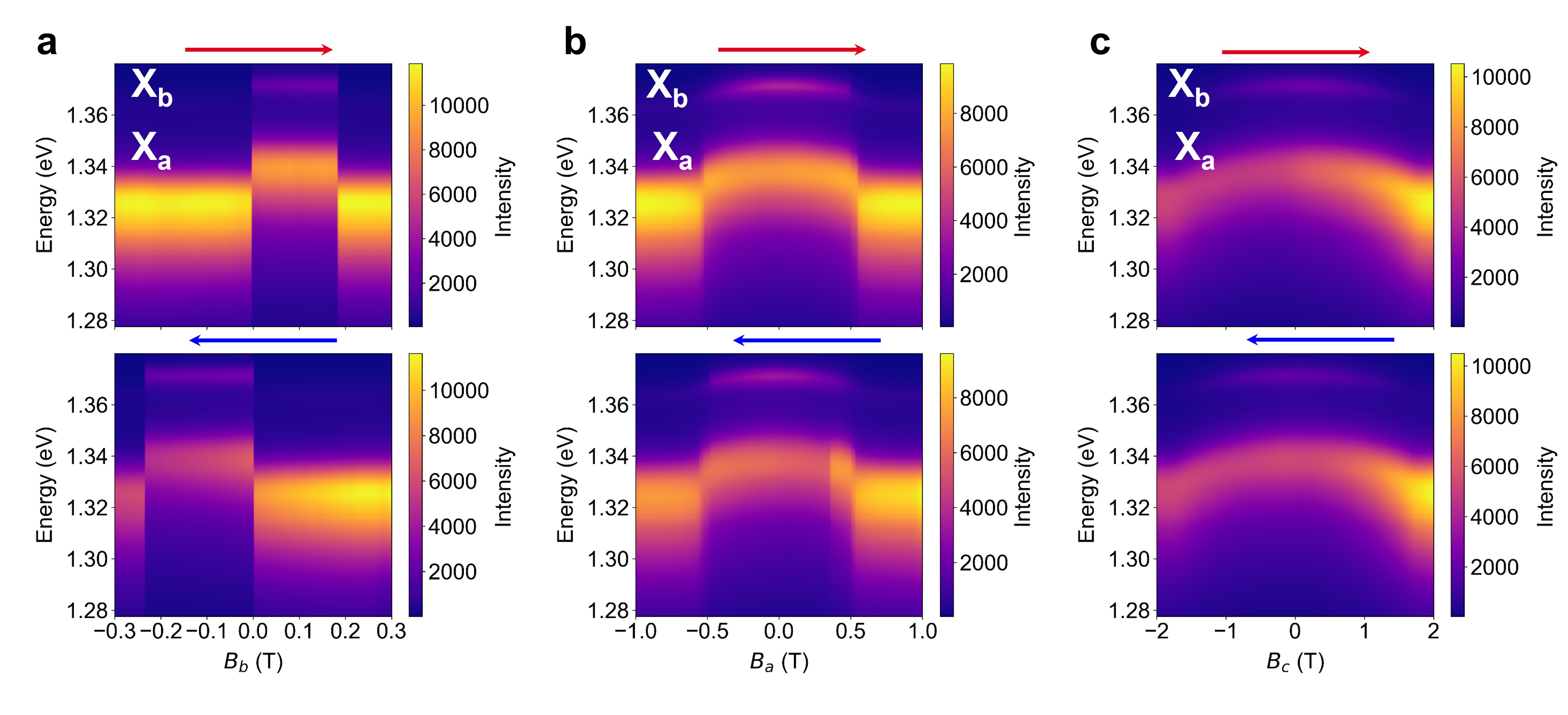}
\caption{\label{fig:figS6}\textbf{Magneto-optical measurements on the twisted bilayer at spot 1 under magnetic fields applied along specific crystallographic directions.}
\textbf{a} shows the magnetic field dependence along the $b$-axis, \textbf{b} corresponds to the $a$-axis, and \textbf{c} represents the $c$-axis.
In each case, the top row for each figure shows data for increasing magnetic field (negative to positive), while the bottom row shows the response for decreasing field (positive to negative).
}
\end{figure*}

In terms of PL intensity, the A and B excitons show contrasting behavior. The B exciton becomes dimmer in the ferromagnetic state, while the A exciton becomes brighter compared to the antiferromagnetic configuration, see axis labels in (Fig.~\ref{fig:figS5}). This trend is consistently observed during magnetic field sweeps along all three principal crystallographic directions (Fig.~\ref{fig:figS6}).

\clearpage
\section{Supplementary Note 5: Evolution and Reproducibility of the hysteresis.}
\label{sec:BmaxHys}

Supplementary Note \ref{sec:BmaxHys} shows the evolution of the magnetic hysteresis curves for the A-exciton as the maximum applied magnetic field ($B_{\text{max}}$) is increased along the $b$-direction. The curves are displayed vertically from bottom to top, corresponding to $B_{\text{max}} = 0$, 0.1, 0.2, and 0.3~T. In each measurement, the magnetic field is swept from –0.3~T up to the designated $B_{\text{max}}$ and then back to –0.3~T, completing a full cycle. This sequential evolution reveals how the hysteresis loop develops with increasing field: the loop gradually widens and its shape changes, directly reflecting the progressive reorientation and modification of the underlying magnetic states within the bilayer. These results highlight the sensitivity of the excitonic magneto-optical response to the strength of the applied magnetic field and show reproducibility of the hysteretic behavior.

\begin{figure*}[htbp]
\includegraphics[width=0.6\linewidth]{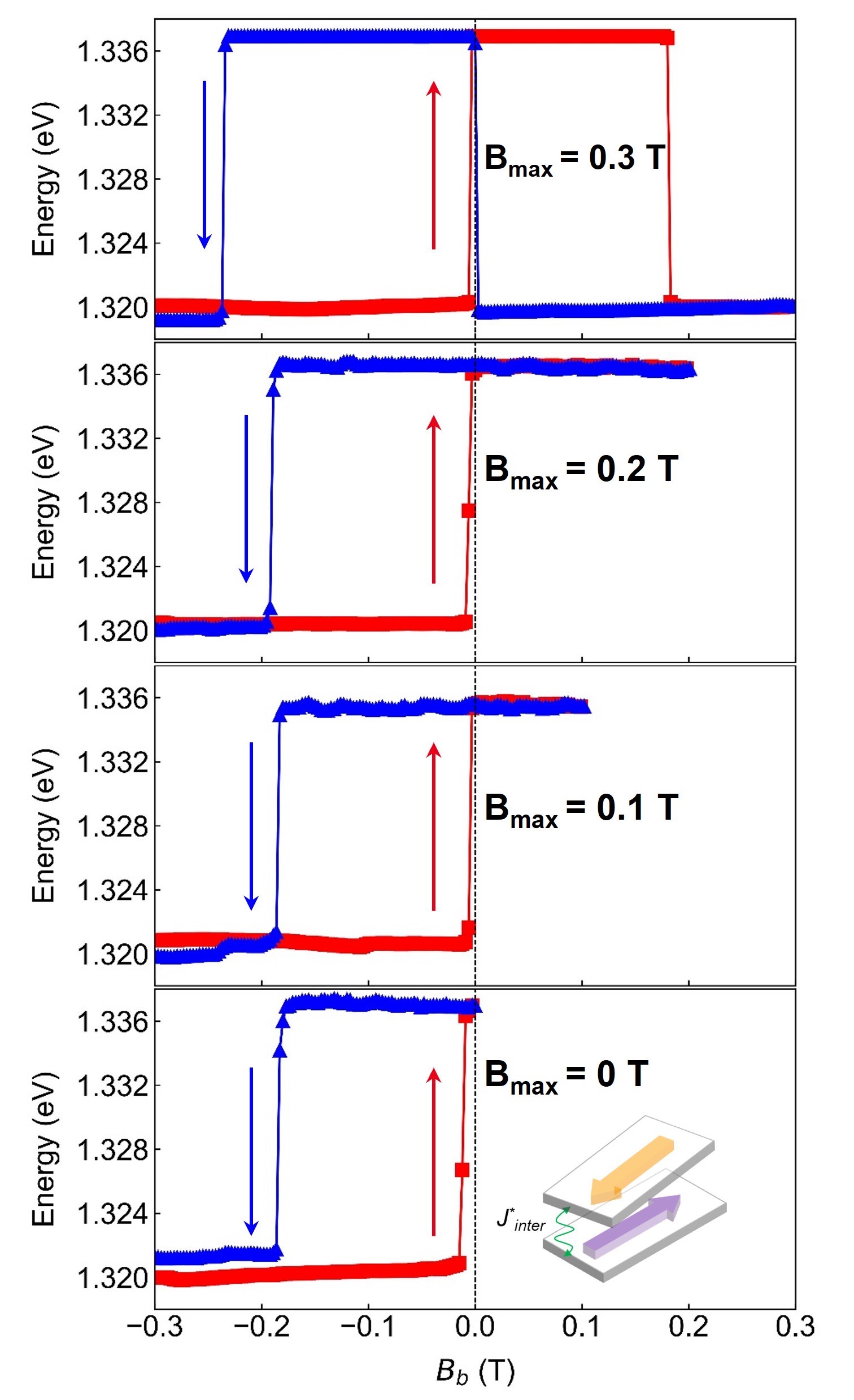}
\caption{\label{fig:figS7}\textbf{Evolution of the magnetic hysteresis curve for the A-exciton with increasing maximum magnetic field ($B_{\text{max}}$) along the $b$-axis.}
The hysteresis curves are displayed vertically in sequential order, starting from the bottom and moving upward, corresponding to maximum fields of $B_{\text{max}} = 0$T, 0.1T, 0.2T, and 0.3T, respectively.
In each case, the magnetic field sweep begins at –0.3T, increases up to the designated $B_{\text{max}}$, and then returns back to –0.3T, completing a closed field cycle.
} 
\end{figure*}

\clearpage
\section{Supplementary Note 6: Hysteresis at different sample positions.}
\label{sec:diffPos}

Supplementary Note \ref{sec:diffPos} presents the magneto-optical response of the twisted bilayer CrSBr discussed in the main text at two distinct locations, spot 2 and spot 3, focusing on the A-exciton. For each spot, the measurements were performed under magnetic fields applied along all the crystallographic directions, highlighting the anisotropic magnetic behavior. The figures show the exciton photoluminescence (PL) response as the magnetic field is swept in both directions. At both spots, a pronounced magnetic hysteresis is observed along the b-axis, while the a- and c-axis directions show no hysteresis. These results demonstrate that the A-exciton PL is highly sensitive to the local magnetic configuration, and that the hysteretic behavior is robust across different regions of the twisted bilayer.

\begin{figure*}[htbp]
\includegraphics[width=0.85\linewidth]{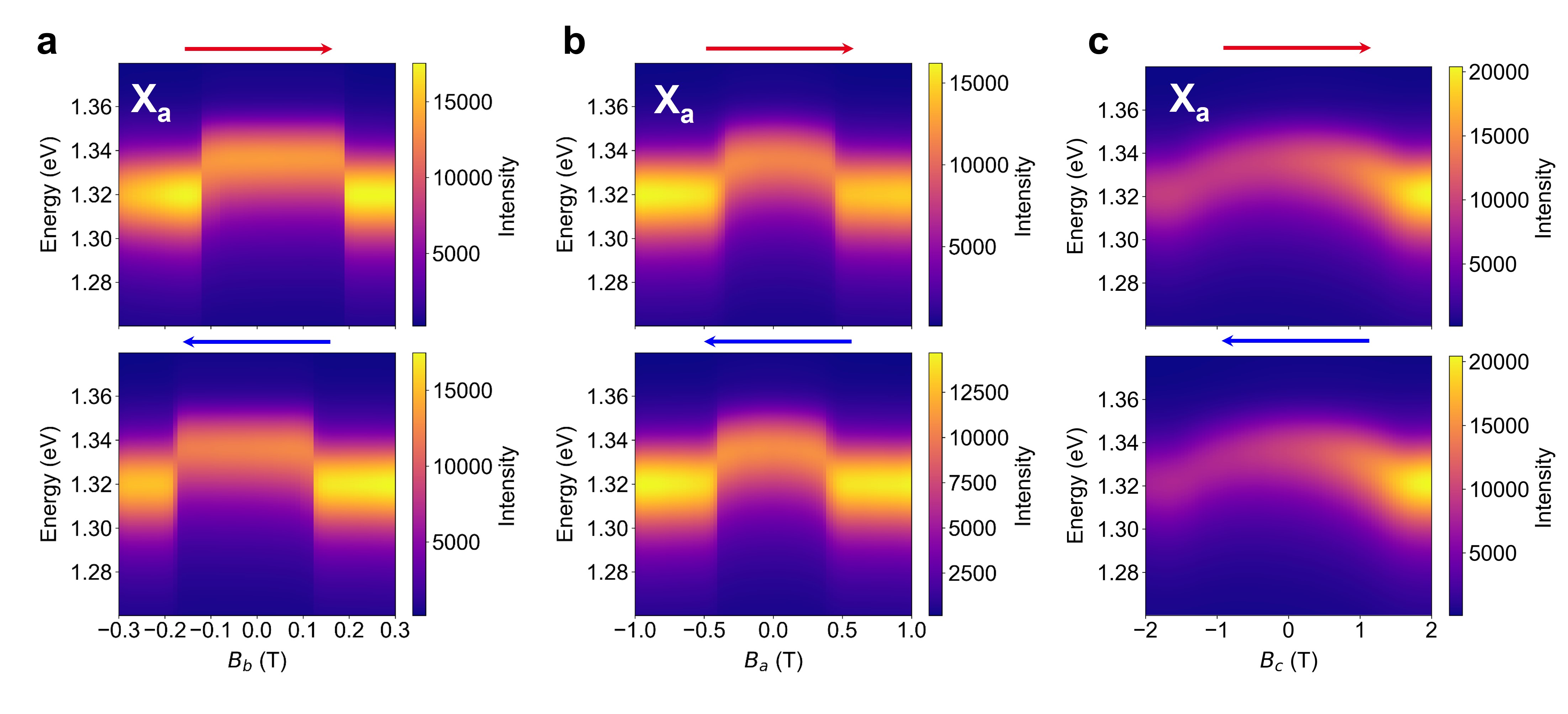}
\caption{\label{fig:figS8} \textbf{Magneto-optical response of the twisted bilayer at spot 2 for the A-exciton, measured under magnetic field applied along specific crystallographic directions.} 
\textbf{a} shows the magneto-photoluminescence (PL) with the field applied along the b-direction, \textbf{b} shows the response for the field along the a-direction, and \textbf{c} corresponds to the c-direction.
In each case, the top row for each fig shows data for increasing magnetic field (negative to positive), while the bottom row shows the response for decreasing field (positive to negative).
 A clear magnetic hysteresis is observed when the field is applied along the $b$-direction, whereas no significant hysteresis is seen for fields applied along the $a$ or $c$ directions. }
\end{figure*}

\begin{figure*}[htbp]
\includegraphics[width=0.85\linewidth]{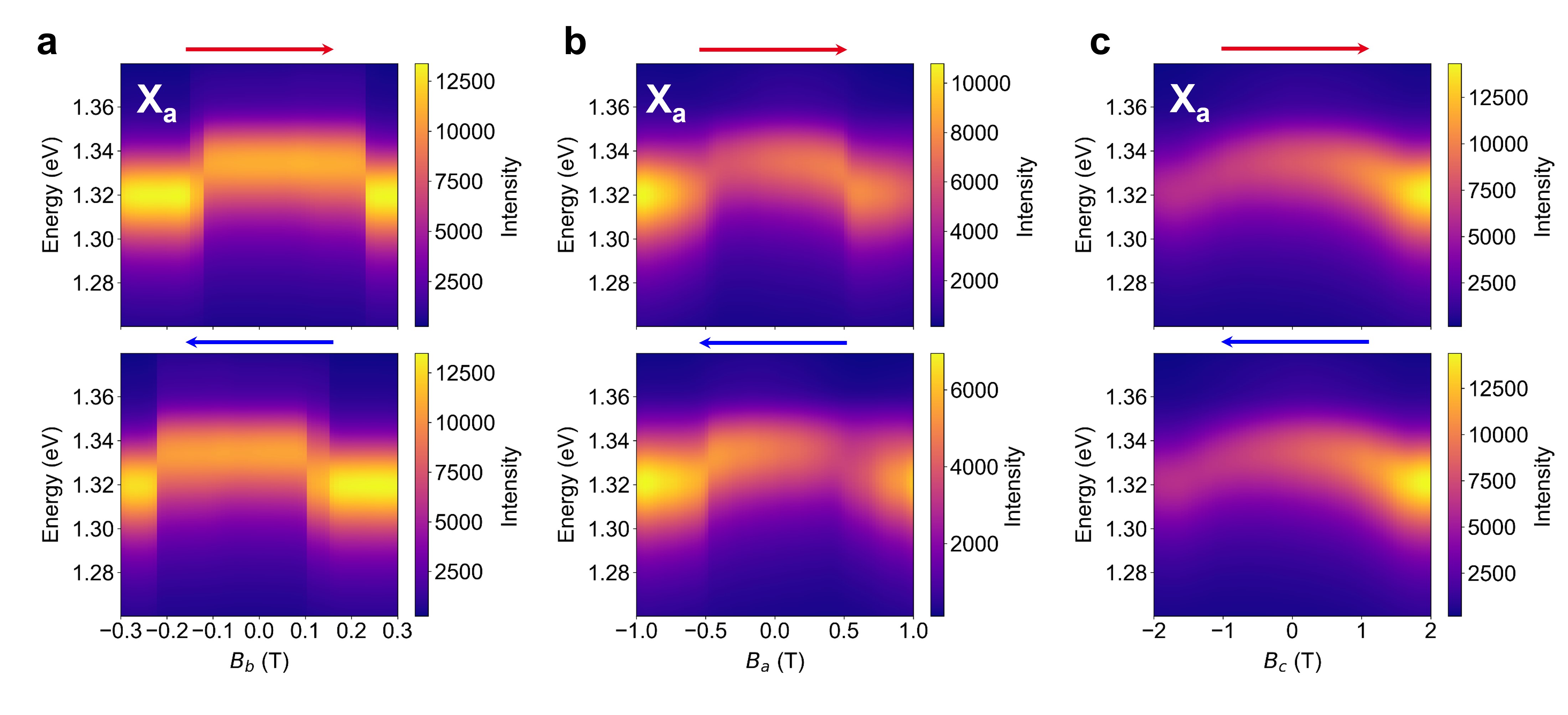}
\caption{\label{fig:figS9} \textbf{Magneto-optical response of the twisted bilayer at spot 3 for the A-exciton, measured under magnetic field applied along three crystallographic directions.}\textbf{a} shows the sample image with the laser spot position for the measurements in dotted circle. 
\textbf{b} shows the magneto-photoluminescence (PL) with the field applied along the b-direction and \textbf{c} corresponds to the c-direction.
In each case, the top row for each fig shows data for increasing magnetic field (negative to positive), while the bottom row shows the response for decreasing field (positive to negative).
 A clear magnetic hysteresis is observed when the field is applied along the b-direction, whereas no significant hysteresis is seen for fields applied along the  c directions. }
\end{figure*}

%

\end{document}